\documentclass[usenatbib]{aastex631}
\usepackage{amsfonts}
\usepackage{amssymb}
\usepackage{amsthm}
\usepackage{amsmath}
\usepackage{alltt}
\usepackage{epstopdf}
\usepackage{graphicx}
\usepackage{float}

\newcommand{\vel}{\boldsymbol{{u}}}
\def\dsdrfig{
\begin{figure}[H]
\centering
\includegraphics[scale=0.3]{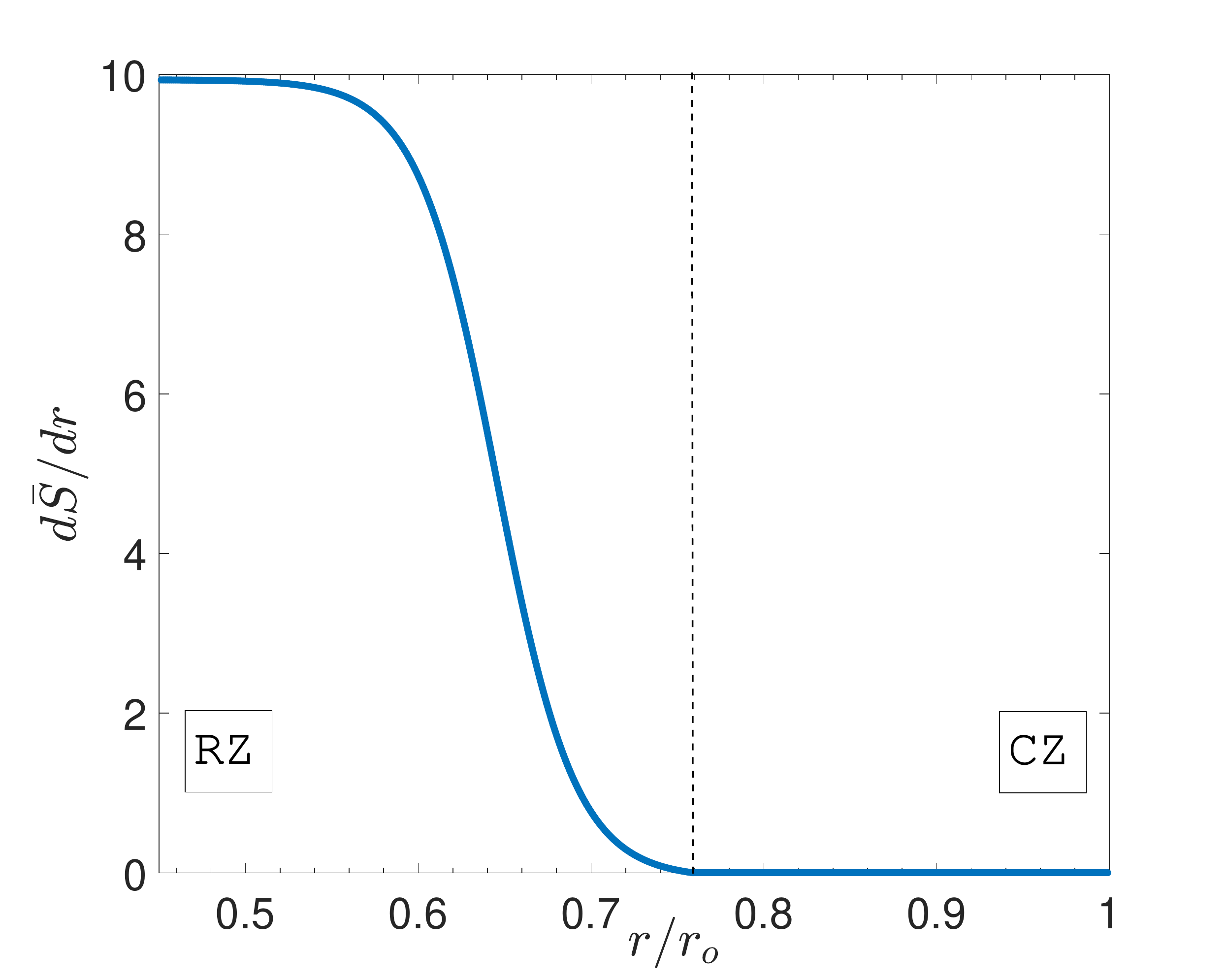}
\caption{Profile of  the non-dimensional background entropy gradient $d\bar{S}/dr$ versus $r/r_o$. The black dashed vertical line corresponds to the base of the convective region.}
\label{fig:dSdr}
\end{figure}
}

\def\Qfig{
\begin{figure}[H]
\centering
\includegraphics[scale=0.3]{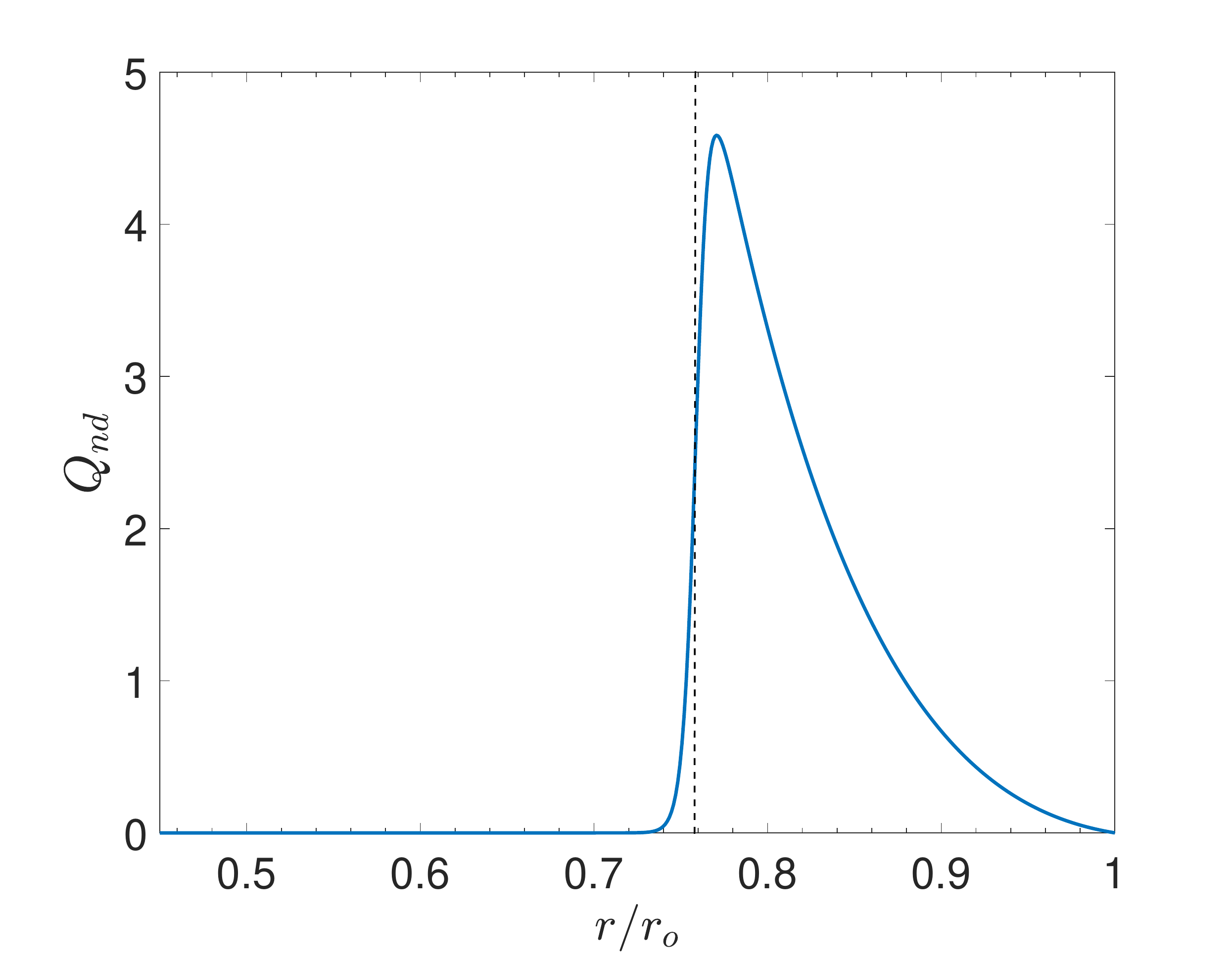}
\caption{Profile of the non-dimensional  heating function $Q_{nd}$ against the radius $r/r_o$ {\color{black}for the case with $N_{\rho}=3$}. The black dashed vertical line corresponds to the base of the convective region.}
\label{fig:Q}
\end{figure}
}

\def\ursNrhofig{
\begin{figure*}[!tbp]
\centering
\includegraphics[scale=0.5]{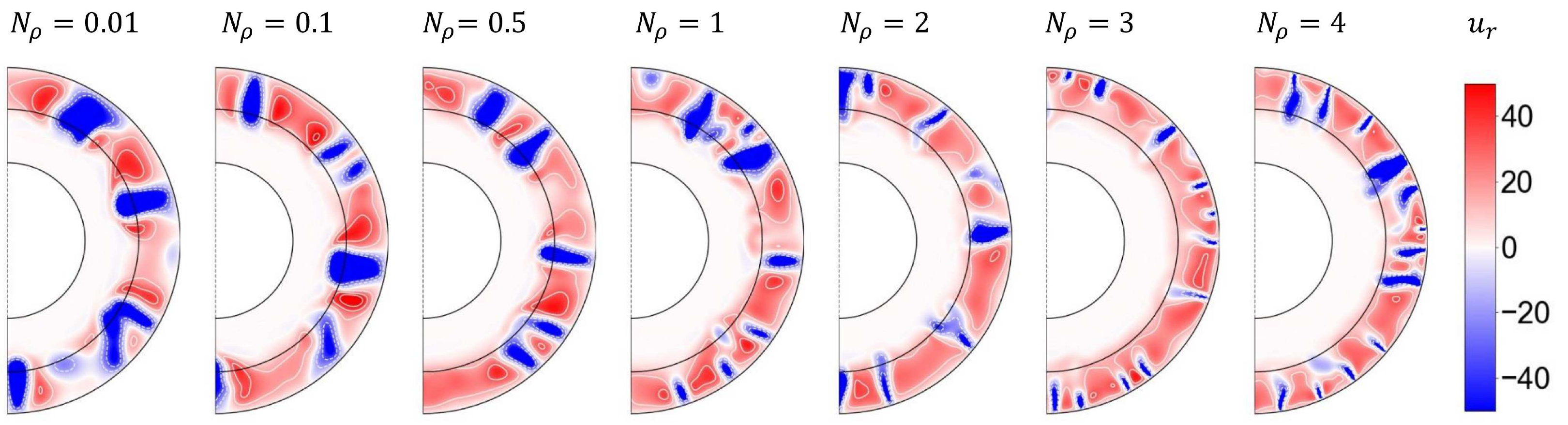}
\caption{Snapshots of meridional slices of the radial velocity $u_r$ at a selected longitude   for different values of $N_{\rho}$ (indicated above each panel) at  Ra $=10^5$ {\color{black} for the non-rotating runs}. The inner black line marks the bottom of the convective region. The convective motions overshoot into the stable region in all $N_{\rho}$ cases and the amount of overshooting depends on $N_{\rho}$.}
\label{fig:uNrho}
\end{figure*}
}

\def\KE_Nrho_allfig{
\begin{figure*}
\centering
\includegraphics[scale=0.3]{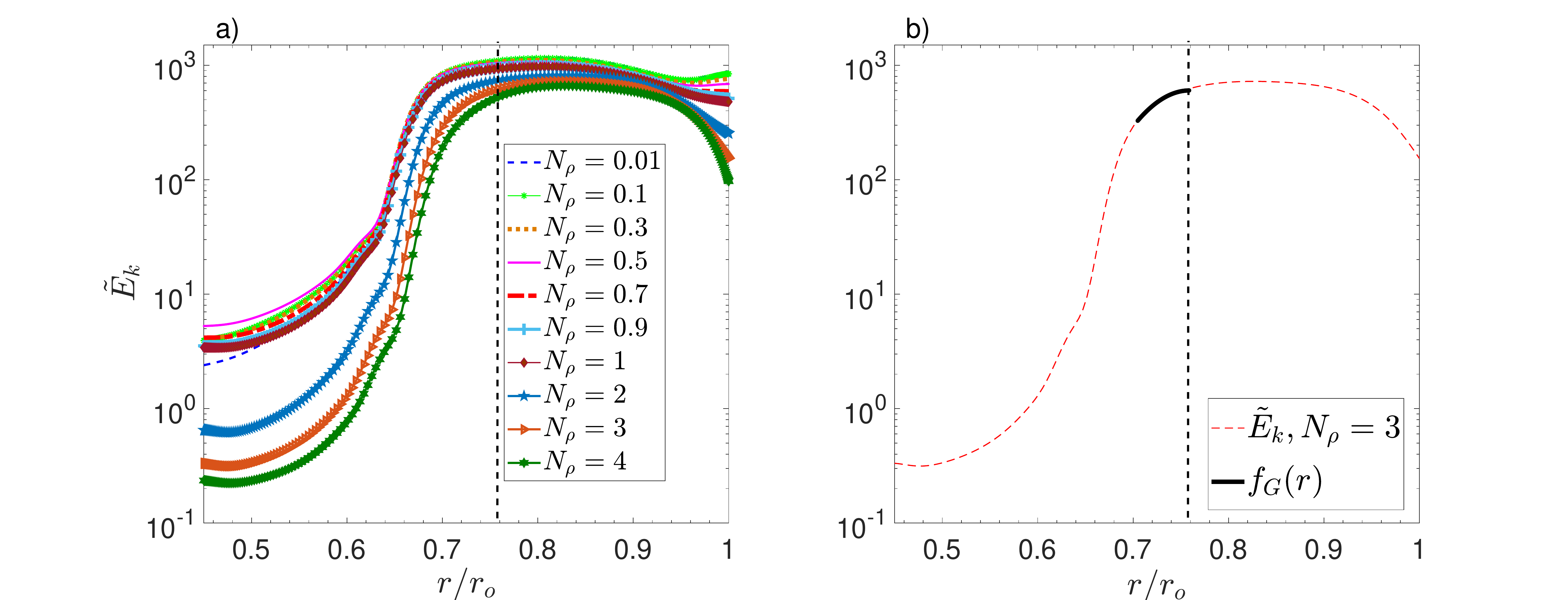}
\caption{Time- and spherically- averaged  kinetic energy profile $\tilde{E}_k(r)$  versus $r/r_o$ at Ra $=10^5$. The black dashed vertical line corresponds to the base of the convective region. a) The profiles of $\tilde{E}_k(r)$ are shown for all of the different $N_{\rho}$ cases. $\tilde{E}_k(r)$ decays as a half-Gaussian below the base of the CZ. b) Profile of $\tilde{E}_k(r)$  along with the   fitted  Gaussian function $f_G(r)$ (see Eq. (\ref{eq:fG})) below the base of the CZ  for the case with $N_{\rho}=3$.}
\label{fig:KE_Nrho_allfig}
\end{figure*}
}

\def\deltaGNrhofig{
\begin{figure}[H]
\centering
\includegraphics[scale=0.3]{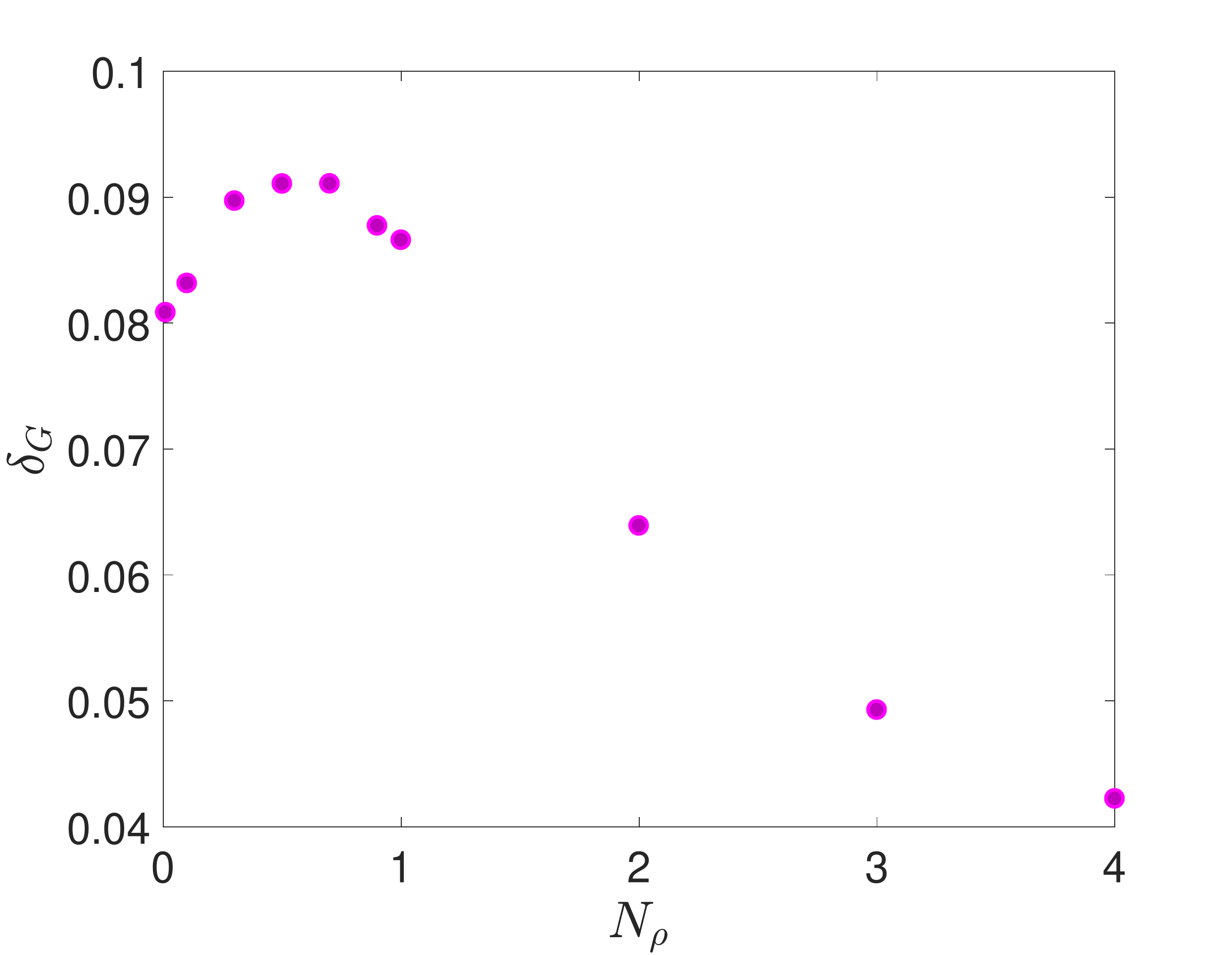}
\caption{Dependence of the computed overshoot lengthscale $\delta_G$ on the  density stratification in the CZ $N_{\rho}$  at Ra $=10^5$. The lengthscale $\delta_G$ increases for $N_{\rho}\lesssim 0.7$ and decreases for $N_{\rho}>0.7$ indicating a non-monotonic dependence on $N_{\rho}$.}
\label{fig:deltaGNrho}
\end{figure}
}

\def\dgRnfig{
\begin{figure*}
\centering
\includegraphics[scale=0.6]{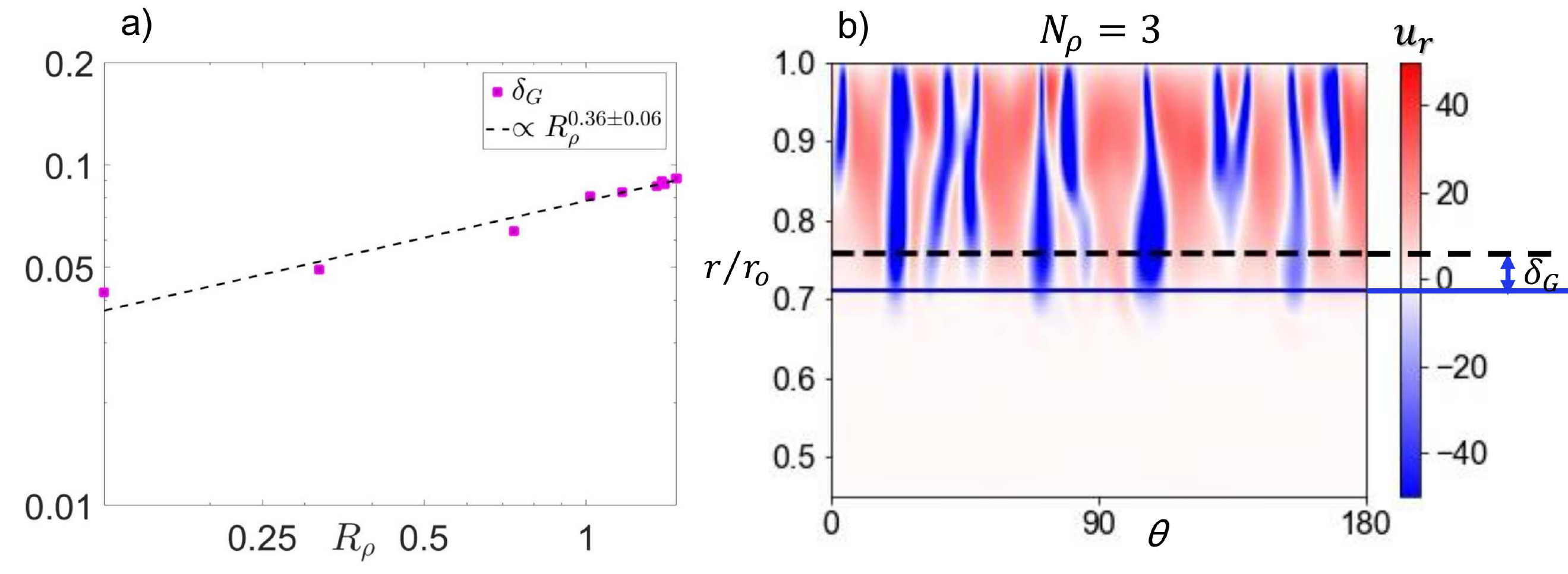}
\caption{a) Dependence of  $\delta_{G}$ on the ratio of the density stratification in the RZ over the one in the CZ  given by the parameter $R_{\rho}$. The overshoot lengthscale {\color{black} is $\delta_G=0.078R_{\rho}^{0.36}$}. b) Snapshot  of the radial velocity $u_r$ in $r/r_o$ and $\theta$ at a fixed longitude  for the Ra $=10^5$ and $N_{\rho}=3$ run. The black dashed line indicates the bottom of the CZ, and the the blue solid line indicates the depth down to $r_c-\delta_G$.  The overshoot lengthscale $\delta_G$ characterizes remarkably well the depth in the RZ down to which the convective motions overshoot on average.}
\label{fig:dgRn}
\end{figure*}
}

\def\dSRafig{
\begin{figure}[H]
\centering
\includegraphics[scale=0.3]{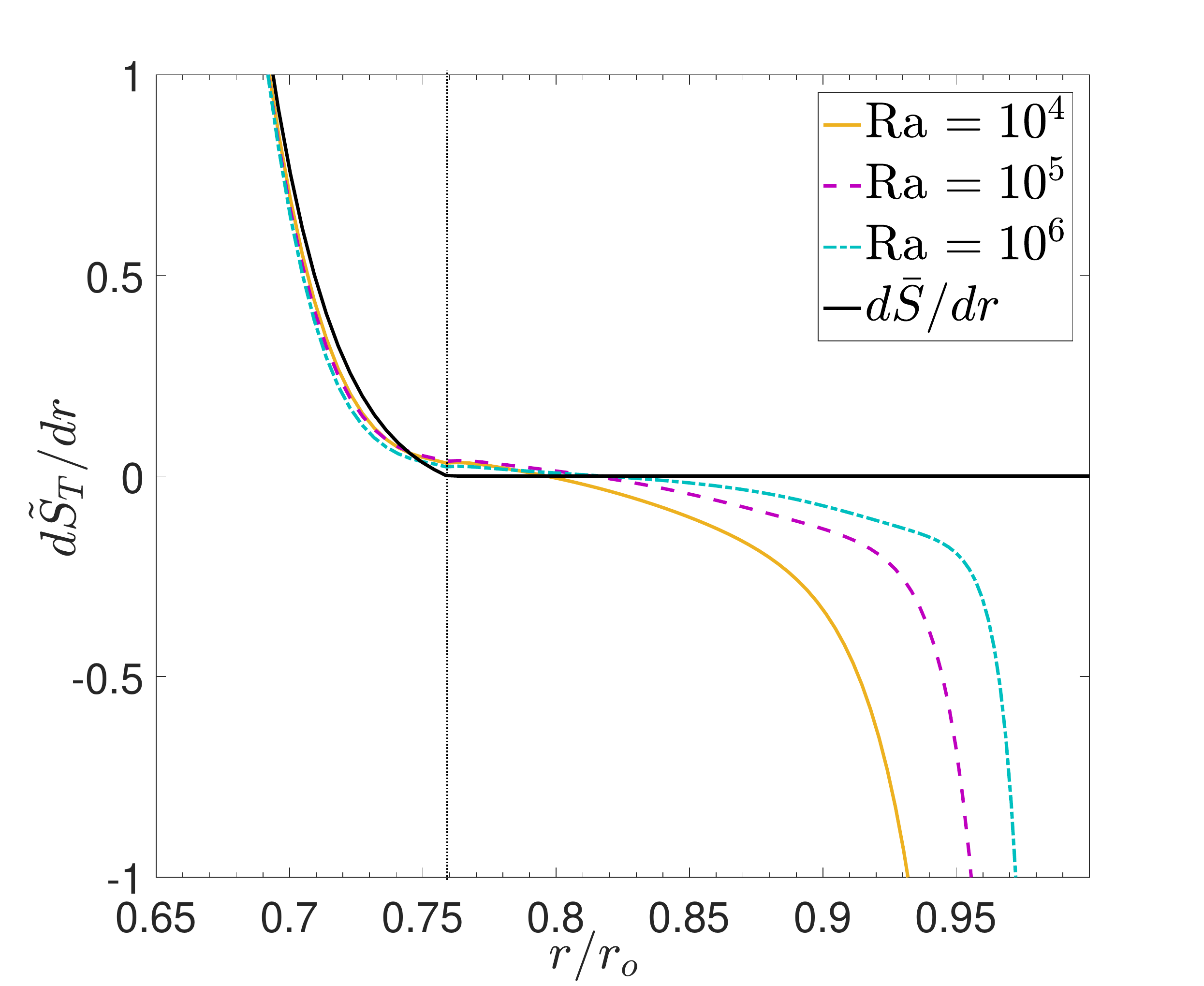}
\caption{Time- and spherically- averaged total adjusted entropy gradient profile 
$d\tilde{S}_T/dr$ against $r/r_o$ at $N_{\rho}=3$ for three different Rayleigh numbers compared with the background $d\bar{S}/dr$. The black  vertical line corresponds to the base of the convective region. Larger values of Ra lead to stronger partial thermal mixing in the RZ and a smaller subadiabatic layer close to the bottom of the CZ. }
\label{fig:dSRa}
\end{figure}
}

\def\dSNrhofig{
\begin{figure}[H]
\centering
\includegraphics[scale=0.3]{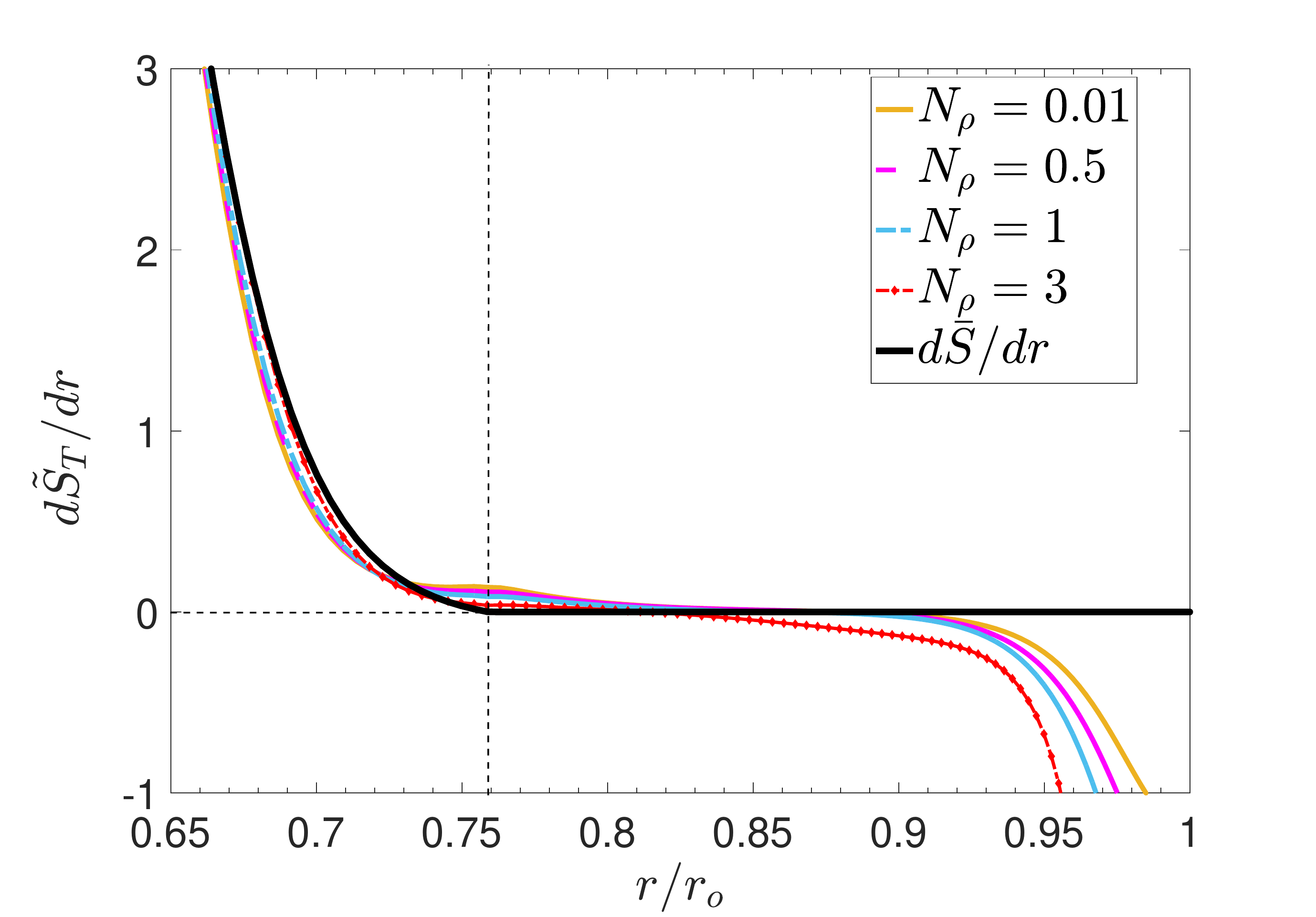}
\caption{Time- and spherically- averaged total adjusted entropy gradient profile $d\tilde{S}_T/dr$ against $r/r_o$ for four $N_{\rho}$ cases compared with the background $d\bar{S}/dr$. The black dashed vertical line corresponds to the base of the convective region. The weak partial thermal mixing in the RZ has a slight dependence on $N_{\rho}$.}
\label{fig:dSNrho}
\end{figure}
}

\def\ur_shell_rotfig{
\begin{figure*}
\centering
\includegraphics[scale=0.5]{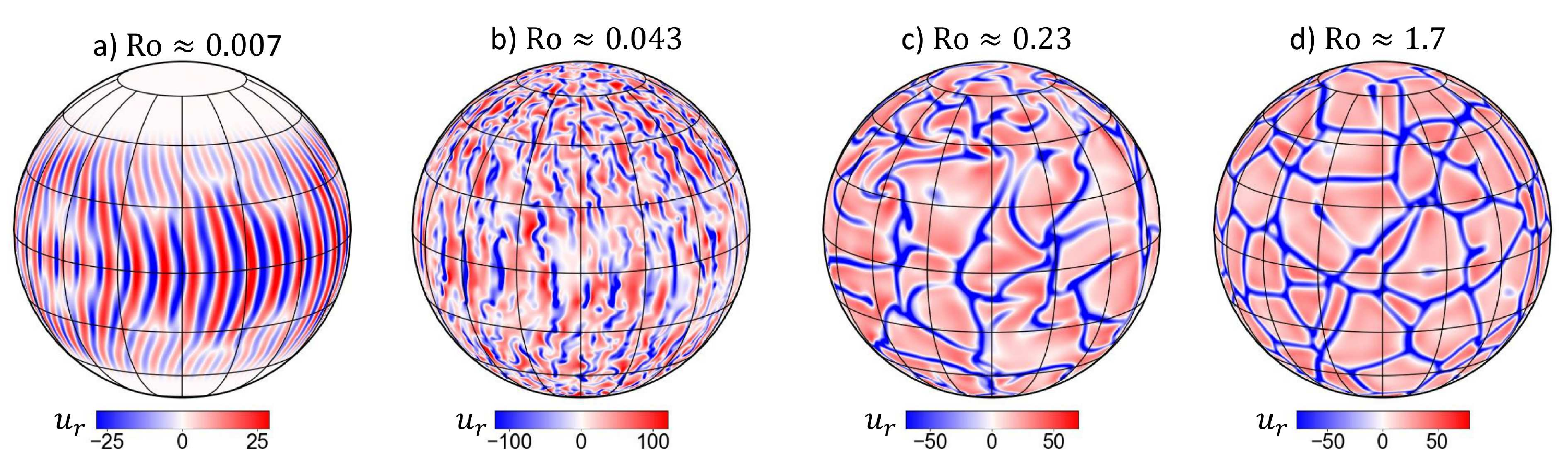}
\caption{Snapshots of shell slices of the radial velocity $u_r$ at $r=0.9r_o$, i.e. close to the surface of the spherical shell for a) Ro $\approx 0.007$ at Ra $=10^5$ and Ek = $0.001$, b) Ro $\approx 0.043$ at Ra $=10^6$ and Ek = $0.001$, c) Ro $\approx 0.23$ at Ra $=10^5$ and Ek = $0.01$, and d) Ro $\approx 1.7$ at Ra $=10^5$ and Ek = $0.1$. The flow becomes more aligned with the axis of rotation with decreasing values of Ro.}
\label{fig:ur_shell_rot}
\end{figure*}
}

\def\Rourfig{
\begin{figure*}
\centering
\includegraphics[scale=0.4]{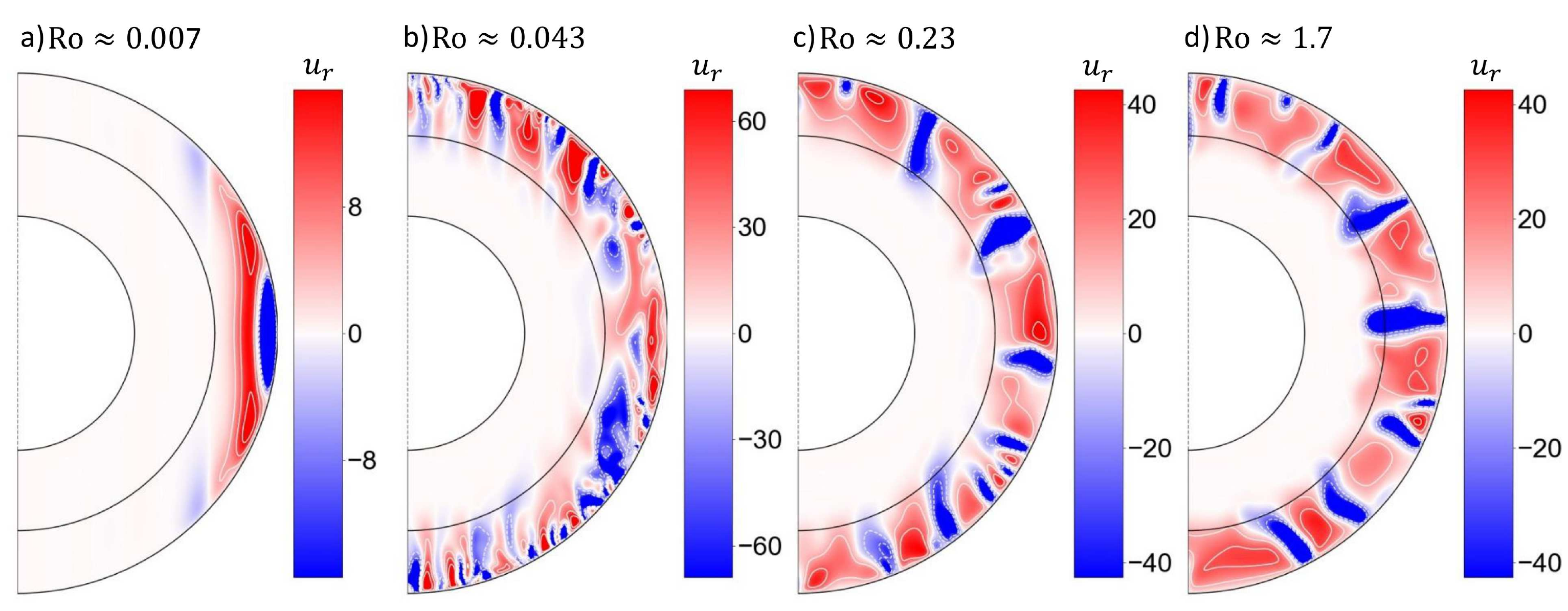}
\caption{Snapshots of meridional slices of the radial velocity $u_r$ varying in $r$ and $\theta$ at a selected longitude for four different values of Ro with a) Ro $\approx 0.007$, Ra $=10^5$ and Ek $=0.001$, b) Ro $\approx 0.043$, Ra $=10^6$ and Ek $=0.001$, c) Ro $\approx 0.23$, Ra $=10^5$ and Ek $=0.01$, and d) Ro $\approx 1.7$, Ra $=10^5$ and Ek $=0.1$. The amount of overshooting below the base of the convection zone seems to be decreasing with decreasing Ro.}
\label{fig:Rour}
\end{figure*}
}

\def\Rofig{
\begin{figure}[H]
\centering
\includegraphics[scale=0.4]{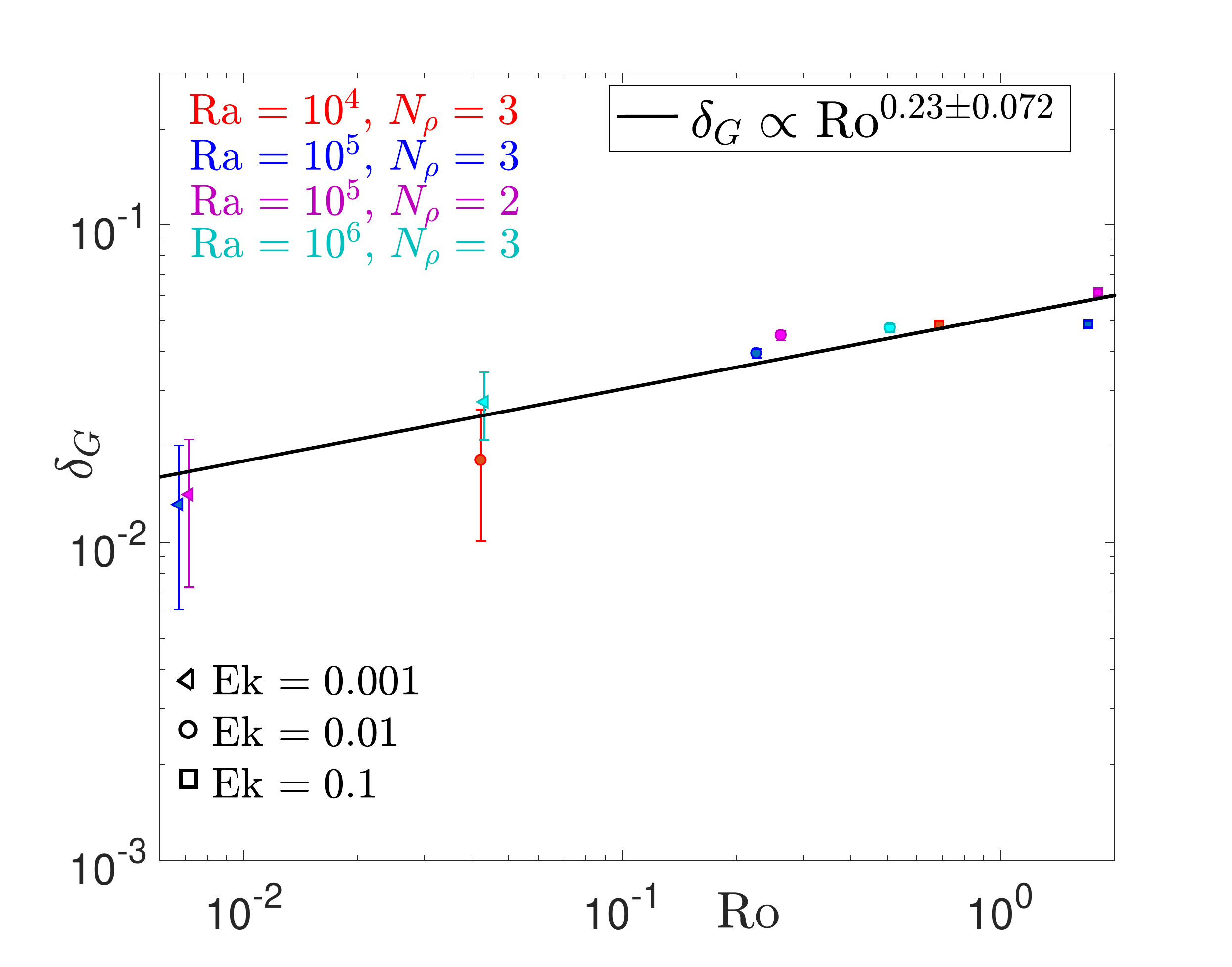}
\caption{Dependence of the computed overshoot lengthscale $\delta_G$ on Rossby number Ro.  Error bars associated with uncertainty in the fit are indicated. The  black line is  the fitted function that describes how  $\delta_G$ scales with respect to Ro, namely {\color{black}$\delta_G\propto{\rm{Ro}}^{0.23\pm 0.072}$}.  The different colors correspond to different Ra and $N_{\rho}$  indicated in the upper left corner of the plot, while the different shapes correspond to the different values of Ek (shown in the lower left corner of the plot).}
\label{fig:Rodov}
\end{figure}
}

\def\deltaGthetafig{
\begin{figure}[H]
\centering
\includegraphics[scale=0.3]{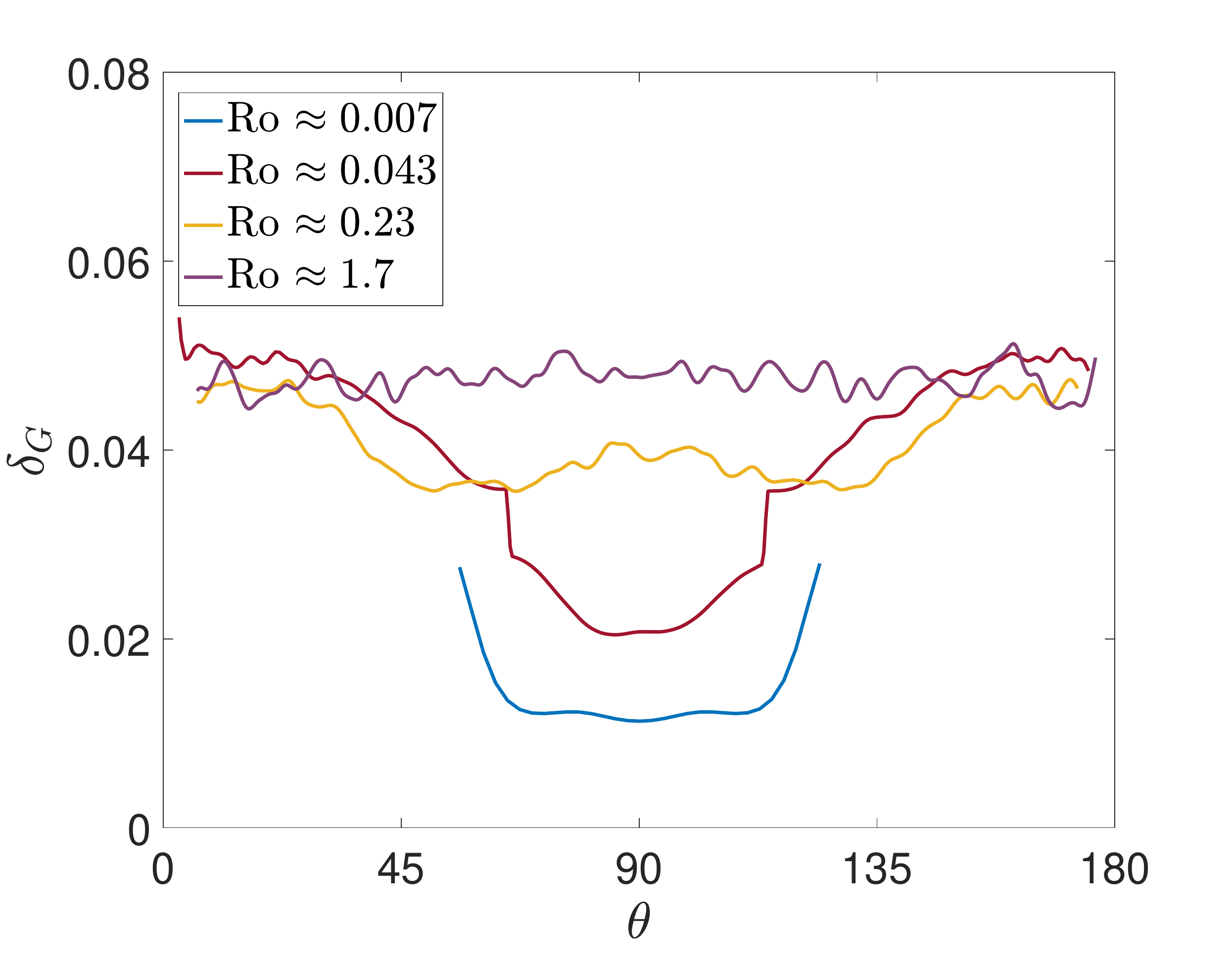}
\caption{Computed overshoot lengthscale $\delta_{G}$ as a function of latitude $\theta$ for four different Ro cases. For larger values of Ro, the latitudinal dependence of $\delta_G$ is weaker.}
\label{fig:deltaGtheta}
\end{figure}
}

\def\Sazfig{
\begin{figure}[!tbp]
\centering
\includegraphics[scale=0.3]{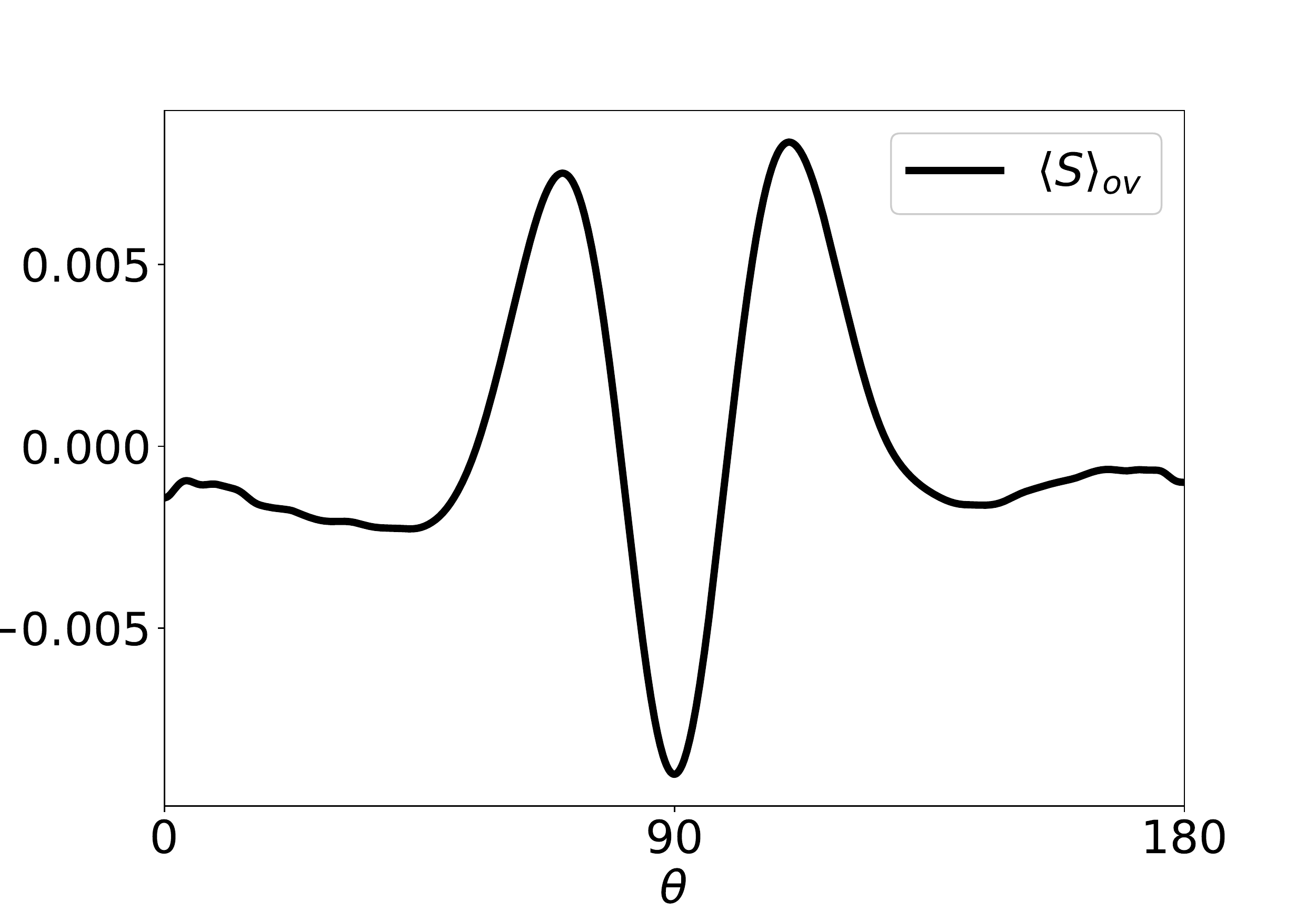}
\caption{Latitudinal dependence of  the time- and azimuthally- averaged entropy perturbations  (with  their spherically symmetric mean subtracted) volume-averaged within the overshoot region  $\langle S\rangle_{ov}$ for the solar-like case with Ro $\approx 0.043$. $\langle S\rangle_{ov}$ does not vary substantially with latitude and exhibits a non-monotonic profile.}
\label{fig:Saz}
\end{figure}
}

\def\FRofig{
\begin{figure*}
\centering
\includegraphics[scale=0.7]{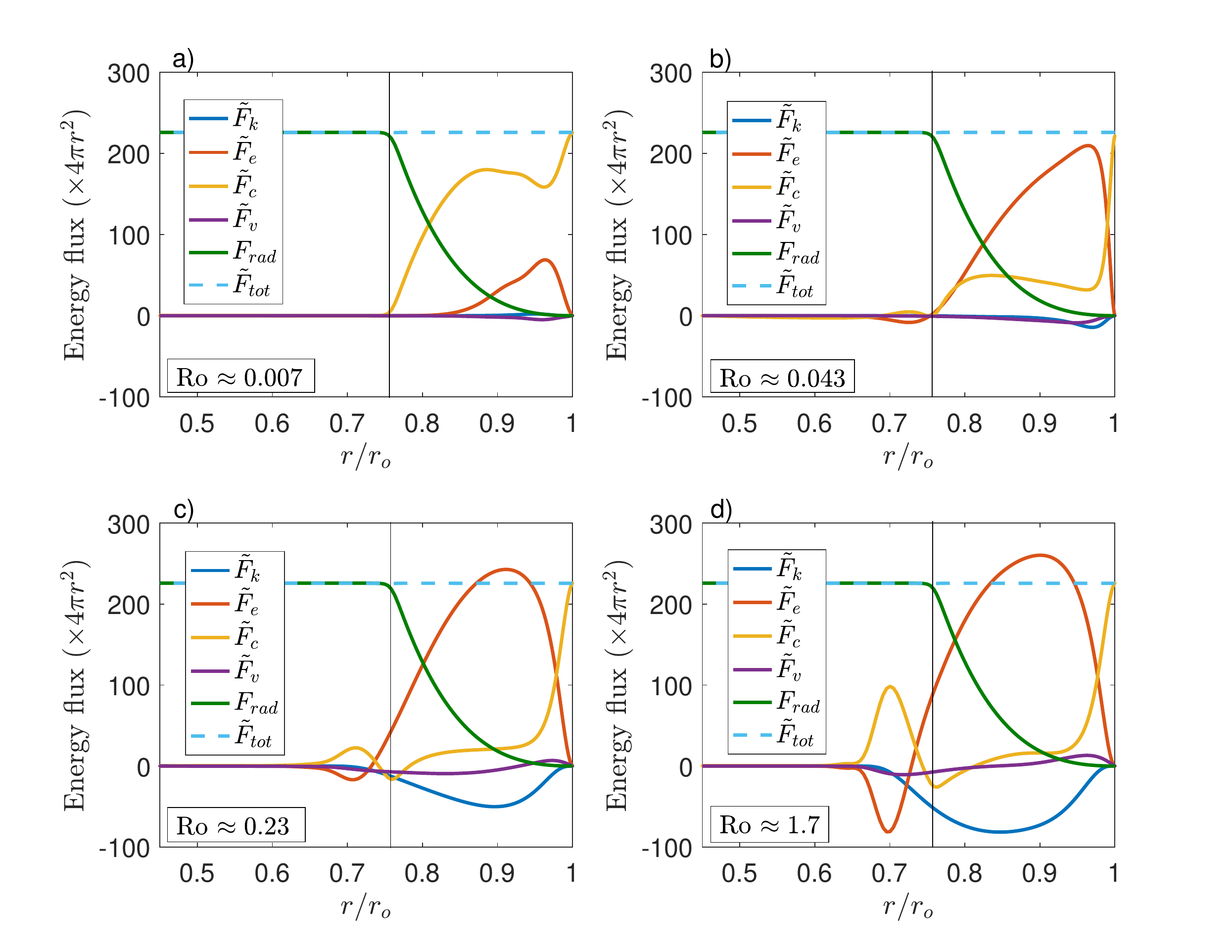}
\caption{Profiles of the time- and spherically- averaged  non-dimensional radial fluxes multiplied by the surface area $4\pi r^2$ against the radius $r/r_o$ with  a) Ro $\approx 0.007$ at Ra $=10^5$ and Ek = $0.001$, b) Ro $\approx 0.043$ at Ra $=10^6$ and Ek = $0.001$, c) Ro $\approx 0.23$ at Ra $=10^5$ and Ek = $0.01$, and d) Ro $\approx 1.7$ at Ra $=10^5$ and Ek = $0.1$. The black  vertical line corresponds to the base of the convective region.}
\label{fig:FRo}
\end{figure*}
}

\def\KEmeanflowsfig{
\begin{figure*}
\centering
\includegraphics[scale=0.3]{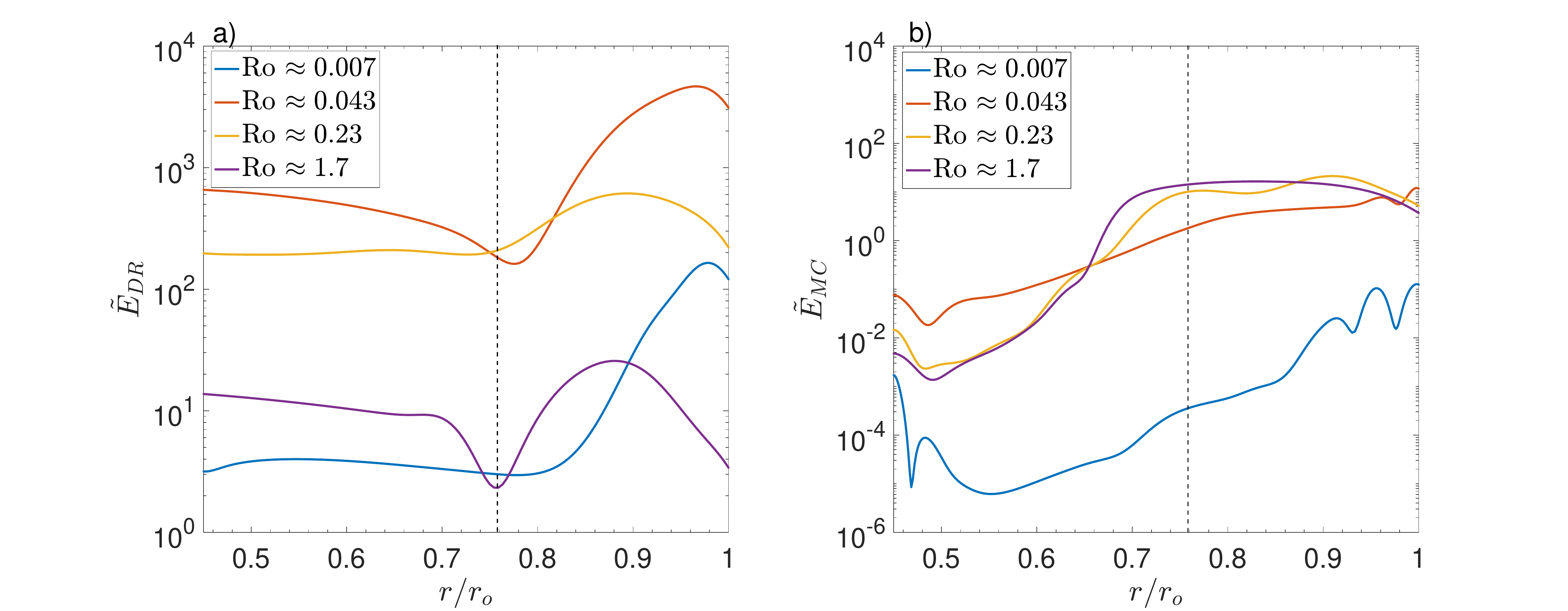}
\caption{Time- and spherically- averaged non-dimensional kinetic energy  in the mean flows. The black dashed vertical line corresponds to the base of the convective region. a) Profile of the kinetic energy   related to the differential rotation $\tilde{E}_{DR}(r)$. $\tilde{E}_{DR}(r)$ is largest for the solar-like and the anti-solar cases in the CZ and the RZ.  b) Profile of the kinetic energy associated with the meridional circulation $\tilde{E}_{MC}(r)$ for four different values of Ro. For the three highest Ro cases, $\tilde{E}_{MC}(r)$ is substantially large below the base of the CZ indicating that the mean flows propagate into the RZ.}
\label{fig:KEmeanflows}
\end{figure*}
}

\def\MCDFRofig{
\begin{figure*}
\centering
\includegraphics[scale=0.6]{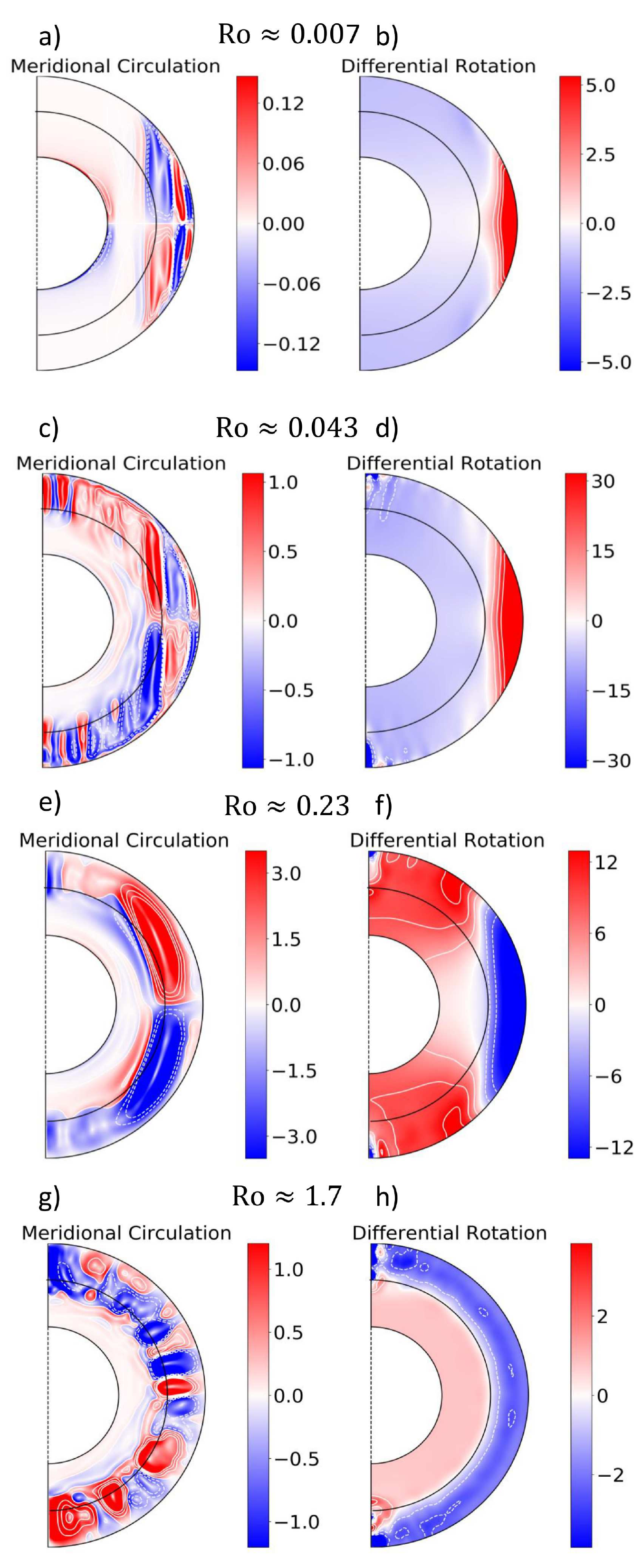}
\caption{Meridional slices illustrating the  meridional circulation streamlines with underlying contours of the mass flux $\pm\sqrt{\langle\bar{\rho}u_r^2\rangle+\langle\bar{\rho}u_{\theta}^2\rangle}$ (blue color corresponds to clockwise motion, and red color to counter-clockwise) and the differential rotation profiles ($\langle u_{\phi}\rangle/(r\sin\theta)$), respectively for a), b)  Ro $\approx 0.007$ at Ra $=10^5$ and Ek = $0.001$, c), d) Ro $\approx 0.043$ at Ra $=10^6$ and Ek = $0.001$, e), f) Ro $\approx 0.23$ at Ra $=10^5$ and Ek = $0.01$, and g), h) Ro $\approx 1.7$ at Ra $=10^5$ and Ek = $0.1$. }
\label{fig:MCDF}
\end{figure*}
}

\def\TWBfig{
\begin{figure*}
\centering
\includegraphics[width=6in,height=4.5in]{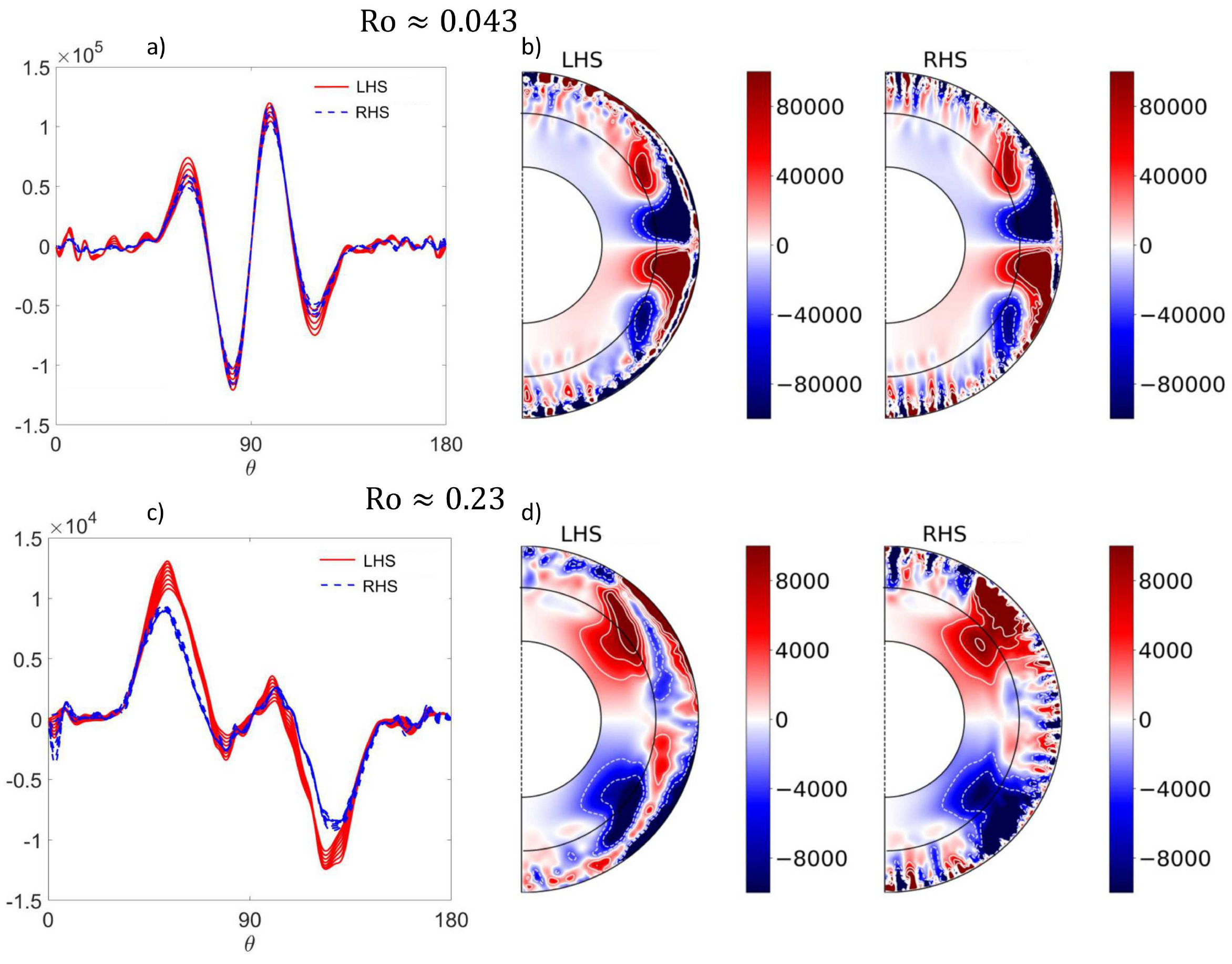}
\caption{Thermal-wind balance for the solar-like and the anti-solar cases. a) Profile of the LHS part and the RHS part of Equation (\ref{eq:TWB2}) versus latitude $\theta$ at different radii within the overshoot region for the solar-like case with Ro $\approx 0.043$ (similarly for panel c) but for the anti-solar case with Ro $\approx 0.23$).  b) Meridional slices of the  LHS and RHS terms illustrating the thermal-wind balance across the whole shell for the solar-like case (similarly for panel d) but for the anti-solar case). The thermal-wind balance is mostly satisfied across the shell for the solar-like case but  is somewhat sustained only in the RZ for the anti-solar case.}
\label{fig:TWB}
\end{figure*}
}

\def\GPRo2fig{
\begin{figure*}
\centering
\includegraphics[scale=0.65]{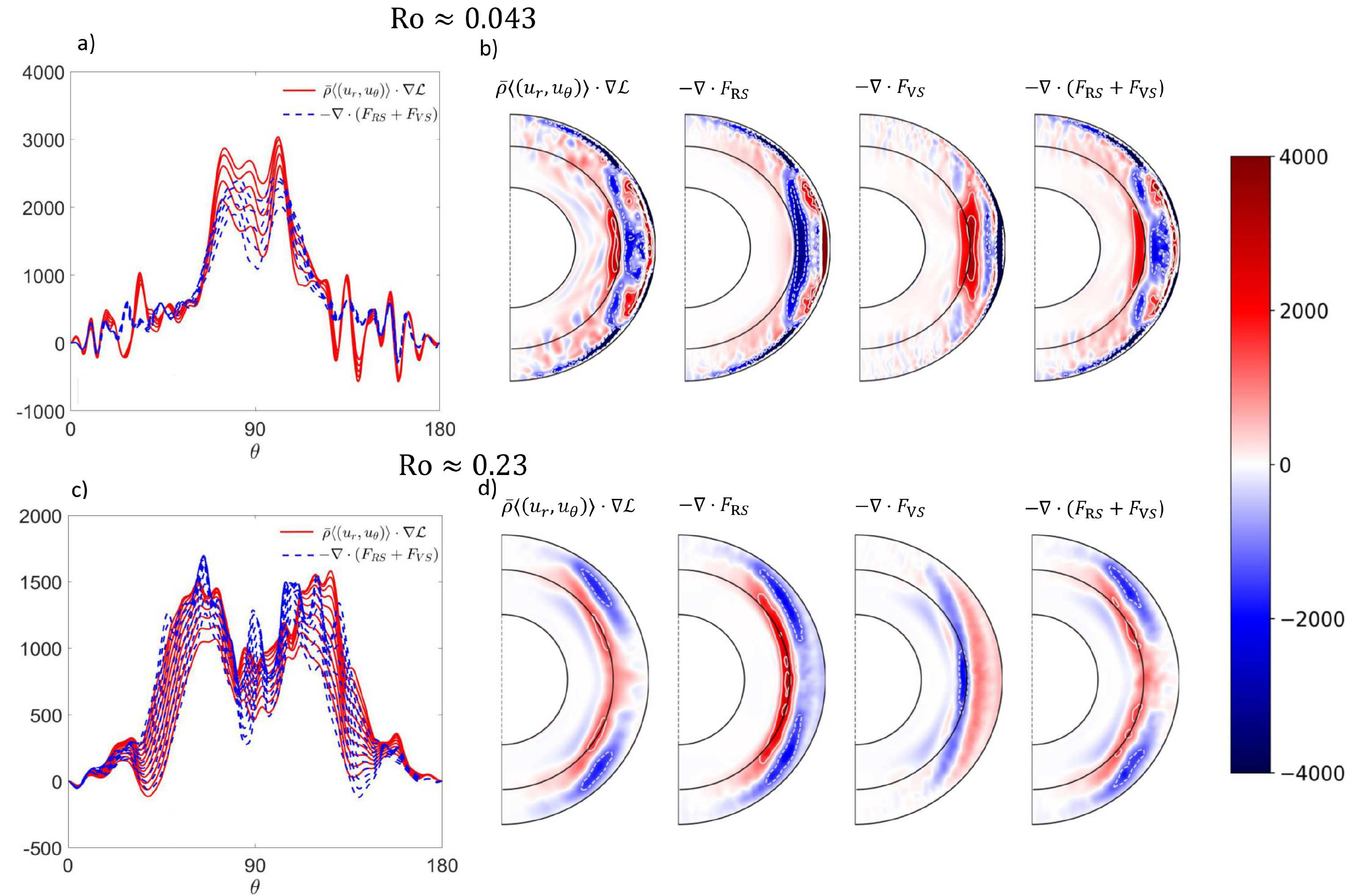}
\caption{Gyroscopic pumping balances for the solar-like and the anti-solar cases. Profile of the LHS and  RHS terms of Equation (\ref{eq:CAM}) versus latitude $\theta$ at different radii within the overshoot region for the solar-like case with Ro $\approx 0.043$ (similarly for panel c) but for the anti-solar case with Ro $\approx 0.23$). b) Meridional slices of the  LHS and RHS terms illustrating the gyroscopic pumping balance across the whole shell for the solar-like case (similarly for panel d) but for the anti-solar case). The gyroscopic pumping balance is overall achieved within the CZ and the overshoot region for both the solar-like case and the anti-solar case.}
\label{fig:GPRo2}
\end{figure*}
}

\def\AMantisolarfig{
\begin{figure*}
\centering
\includegraphics[scale=0.5]{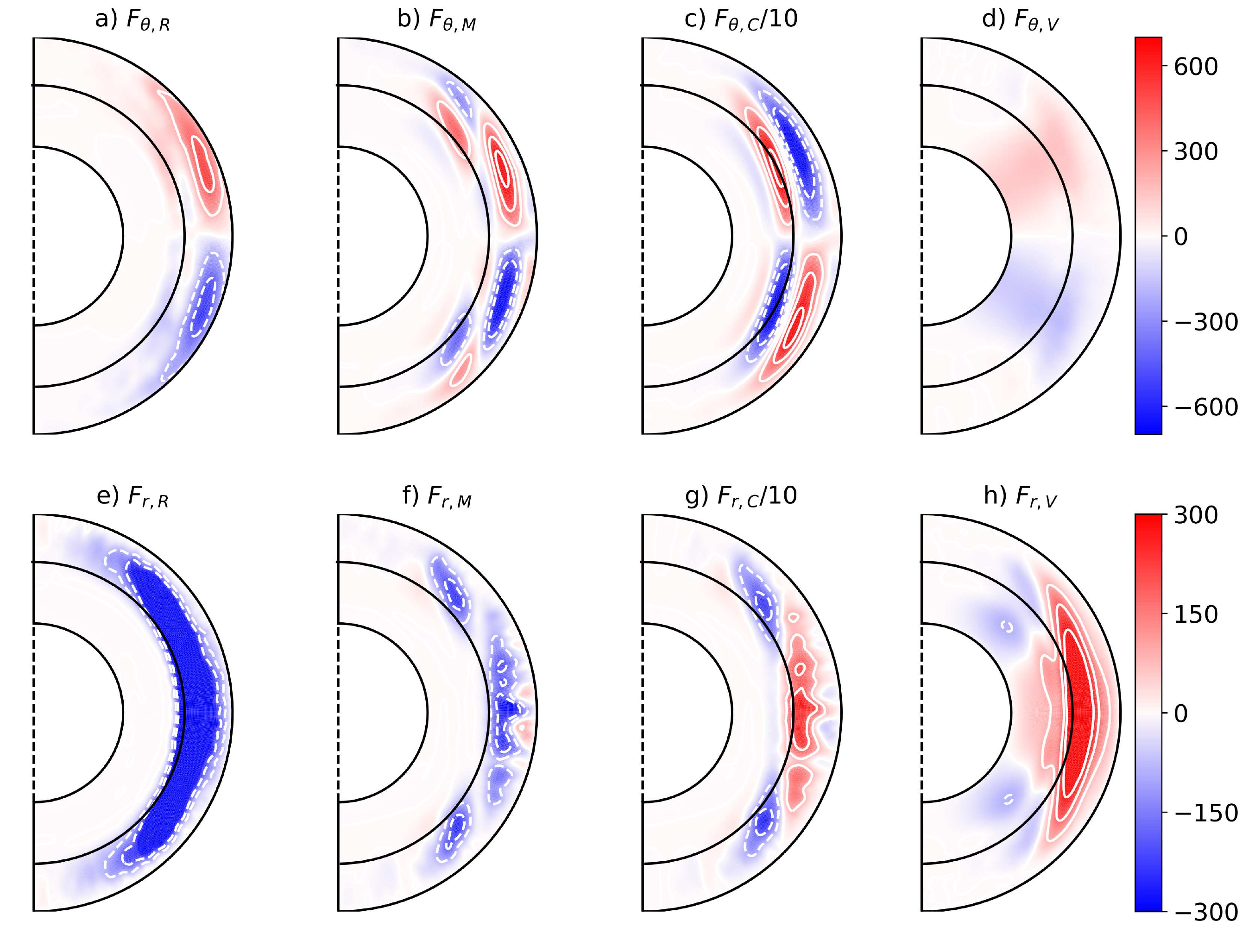}
\caption{Time- and azimuthally- averaged latitudinal and radial fluxes contributing to the angular momentum transport for the anti-solar case with Ro $\approx 0.23$. The fluxes associated with the Reynolds stresses are shown in panels a) and e), the ones related to the mean flows are shown in b) and f), the ones related to the Coriolis term are shown in c) and g) and the ones associated with the viscous stresses are shown in panels d) and h). The fluxes corresponding to the action of the Coriolis force on the mean flows are dominant and as such we have divided them by 10 (panel c) and g)) in order to highlight the profiles of the other weaker fluxes and  examine all of them under the same colorbar.}
\label{fig:AMantisolar}
\end{figure*}
}

\def\AMsolarfig{
\begin{figure*}
\centering
\includegraphics[scale=0.5]{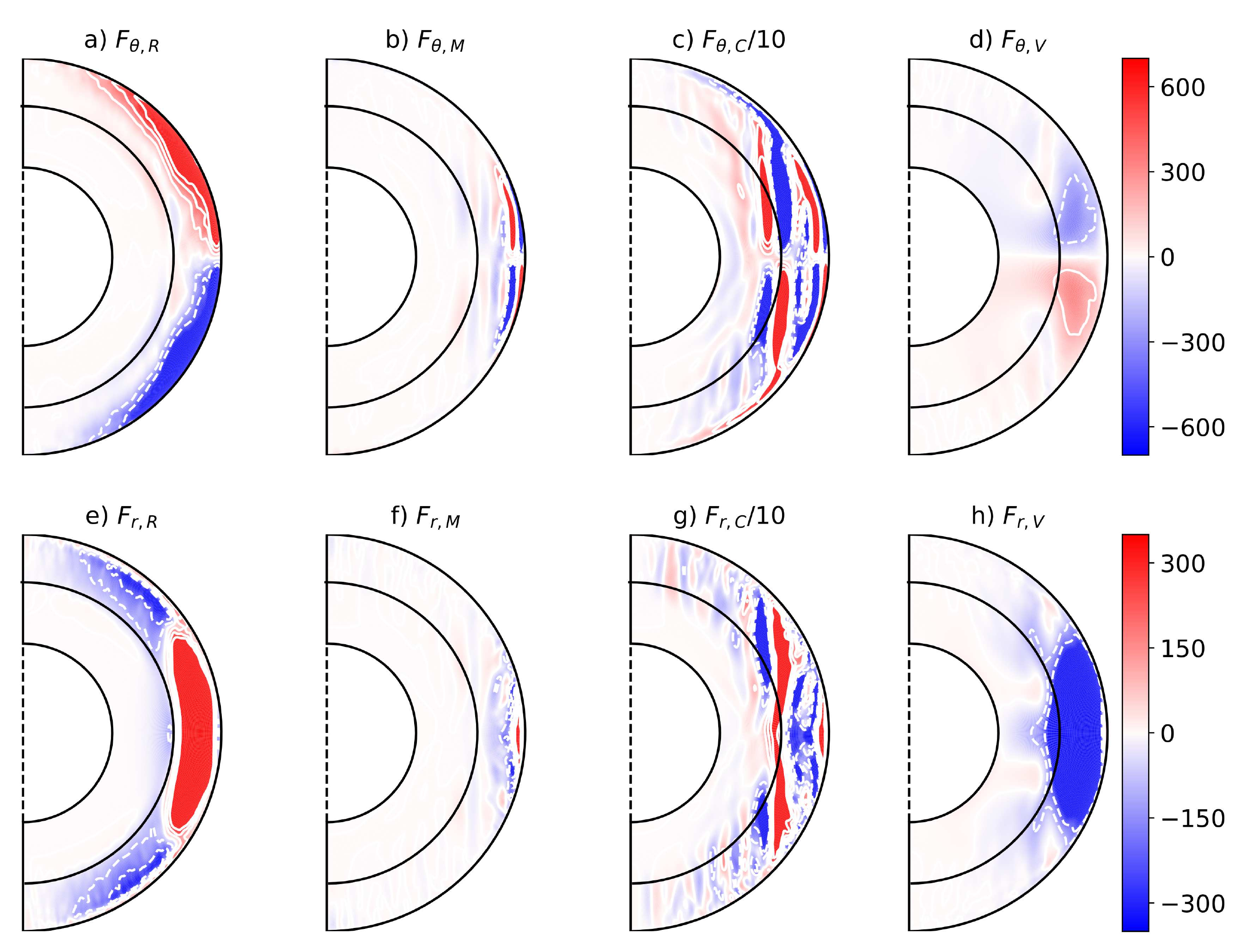}
\caption{Time- and azimuthally- averaged  latitudinal and radial fluxes contributing to the angular momentum transport for the solar-like case with Ro $\approx 0.043$. The fluxes associated with the Reynolds stresses are shown in panels a) and e), the ones related to the mean flows are shown in b) and f), the ones related to the Coriolis term are shown in c) and g) and the ones associated with the viscous stresses are shown in panels d) and h). The fluxes corresponding to the action of the Coriolis force on the mean flows are dominant and as such we have divided them by 10 (panel c) and g)) in order to highlight the profiles of the other weaker fluxes and  examine all of them under the same colorbar.}
\label{fig:AMsolar}
\end{figure*}
}

\begin{document}
\title{On the dynamics of  overshooting convection in spherical shells: Effect of density stratification and rotation}

\correspondingauthor{Lydia Korre}
\email{lydia.korre@lasp.colorado.edu}

\author[0000-0002-0963-4881]{Lydia Korre}
\affiliation{Laboratory for Atmospheric and Space Physics, Boulder, CO 80303, USA}

\author{Nicholas A. Featherstone}
\affiliation{Southwest Research Institute, Department of the Space Studies, Boulder, CO 80302, USA}


\begin{abstract}
Overshooting of  turbulent motions from  convective regions into adjacent stably stratified  zones plays a significant role in stellar interior dynamics as this process may lead  to mixing of chemical species, and contribute to the transport of  angular momentum and magnetic fields.
We present a series of fully non-linear, three-dimensional (3D)   anelastic simulations of overshooting convection in a spherical shell which are focused on the dependence of the overshooting dynamics on the density stratification  and the rotation, both key ingredients in stars which however have not been studied systematically  together via  global simulations. We demonstrate that the overshoot lengthscale  is not simply a monotonic function of the density stratification in the convective region but  instead, it depends on the ratio of the density stratifications in the two zones. Additionally, we find that the overshoot lengthscale decreases with decreasing Rossby number Ro and scales as Ro$^{0.23}$ while it also depends on latitude  with higher Rossby cases leading to a weaker latitudinal variation. We  examine the mean flows arising due to rotation and find that they extend beyond the base of the convection zone into the stable region.  
Our findings may provide a better understanding of the dynamical interaction between stellar convective and radiative regions, and motivate future studies particularly related to the solar tachocline and the implications of its  overlapping with the overshoot region.

\end{abstract}

\section{INTRODUCTION}
\label{sec:intro}
The dynamical interaction  of  convectively unstable zones (CZ) with  stably stratified radiative zones (RZ)  has been a crucial yet still poorly understood problem in the astrophysical context, despite  the fact that such regions exist side-by-side in almost every star and across  many  evolutionary phases.  The boundaries between such stellar regions are not impenetrable, allowing turbulent convective motions generated in the convectively unstable layer to travel into the adjacent stable regions through inertia. This process, termed ``overshooting", has substantial implications for stellar evolution as it can lead to changes in stellar surface element abundances and differential rotation through the  mixing of chemical species and the transport of angular momentum \citep[see e.g.,][]{Pinsonneault,Baraffe17}.  It may also alter the thermal stratification of the radiative zone, converting a portion of it into an adiabatic region that forms an extended CZ, in which case the process is usually termed ``penetration" \citep{Zahn1982,Zahn91}.  Furthermore, overshooting can advect magnetic fields from the CZ into the RZ via turbulent pumping \citep[see e.g.,][]{Tobias2001,Korre21},  {\color{black} and may lead to the establishment of an interface dynamo at the base of the solar convection zone} \citep{Parker93,Charbonneau97}.
\\

\newpage

\subsection{Previous Studies of Non-rotating Overshooting/Penetrative Convection}

\par There exists a large body of astrophysical literature on the topic of convective overshooting.  Early studies were performed using modal expansions \citep[e.g.,][]{LSTZ76,TZLS76,Latour81} or within the context of mixing-length theory \citep[see, for instance][]{ShavivSalpeter1973,Maeder75}.  However, as discussed in \citet{Renzini},  there were many inconsistencies among the results of the mixing-length theory studies due to their sensitivity to the underlying model assumptions. As a result, convective overshooting research came to rely on  alternative, more sophisticated approaches, ranging from early modal-expansion models to modern, fully non-linear computational models. These studies have been mainly focused on (1) the variation of the lengthscales associated with the overshooting motions in response to stellar parameters/properties and (2) the degree to which overshooting convection modifies  the stable background stratification (thermal mixing) and the parameters under which it manages to turn part of it into  an adiabatic layer (penetrative convection).  

Early work focused on penetrative convection was performed  by \citet{Zahn1982}, who employed a modal expansion approach to study overshooting in a Boussinseq system and found substantial penetration within the stable region.  The effects of background density stratification were later included in the study of \citet{Mass84} who demonstrated that increasing stratification in the CZ may lead to larger penetration lengthscales due to the enhanced flow asymmetries arising from compressibility effects.  Scaling arguments have also been applied, as in \citet{Zahn91} who identified two dynamical regimes of interest below the base of the convective region:  a penetrative layer for high P\'{e}clet numbers and a thermal adjustment (overshoot) layer for low P\'{e}clet numbers (where the P\'{e}clet number characterizes the strength of thermal advection relative to thermal diffusion).   The existence of these two regimes was subsequently confirmed through the application of 2D, fully non-linear simulations carried out in Cartesian and cylindrical geometries  \citep[e.g.,][]{Hurlburt86,Hurlburt94, Rogers2005, Rogers2006}.

One of the first systematic studies of overshooting/penetrative convection  via fully non-linear  simulations was that of \citet{Hurlburt94} \citep[also see][]{Hurlburt86} who  performed 2D Cartesian compressible calculations.  That study explored the dependence of overshooting on the so-called stiffness parameter, a measure of the  degree of relative stability of the RZ compared with the CZ.  It also identified  two scaling regimes, consistent with the findings of \citet{Zahn91}, and showed that the penetrative regime was associated with lower stiffness parameters and larger penetration depths while  the overshoot regime was related to higher stiffness parameters and possessed smaller overshoot lengthscales.  

These results have largely been confirmed within the context of fully non-linear 3D simulations.  Studies carried out in both Cartesian and spherical geometries have demonstrated that the overshoot lengthscale decreases with increasing stiffness  {\color{black}parameter} \citep{Brummell, Korre19}. However, as noted in \citet{Korre19}, the transition width between the CZ and the RZ also appears to play an important role in determining the amount of overshooting below the base of the convective region.  Interestingly, these 3D studies did not identify pure penetration, namely the extension of the CZ further down into the stable RZ, but instead, they only observed partial thermal mixing in the RZ, associated with the stronger downflows.   

This lack of pure penetration may be attributed to the lower filling factor of the plumes for 3D simulations relative to 2D simulations when run in otherwise similar parameter regimes \citep[for more on this, see e.g.,][]{Brummell, Rempel2004}.  It may also result from the tendency of the overshoot lengthscale and degree of penetration to be established by strongest downflows \citep[see e.g.,][]{Pratt17,Korre19}. Both the strength of these downflows and the  P\'{e}clet numbers achieved in simulations are explicitly dependent on the degree of thermal forcing as expressed through a Rayleigh number.  The dependence of the overshoot lengthscale on the Rayleigh number has been explored in several studies \citep[see][]{Brummell,Rogers2005,Korre19}, but different scalings were observed, possibly due to the different set-up configurations, geometries, and boundary conditions.  Large density stratification, which enhances the asymmetry between upflows and downflows, may also lead to stronger overshooting and eventually an adiabatic penetrative layer within the RZ.  This effect was discussed in \citet{Mass84}, but has yet to be investigated systematically via fully non-linear, 3D global simulations.

\subsection{The Effects of Rotation}
Relatively few studies have explicitly examined the response of overshooting to the presence of rotation.  While the details vary, these studies suggest that the overshooting depth $\delta$ decreases with decreasing Rossby number Ro (where Ro expresses the relative strength of inertial to Coriolis forces).  Cartesian studies carried out in a rotating {\color{black}$f$}-plane by \citet{Brummell} found that, for a limited range of Ro, $\delta\propto$ Ro$^{0.15}$. Similar results were reported by \citet{Pal2007} who found that $\delta\propto$ Ro$^{0.2}$ at the poles and $\delta\propto$ Ro$^{0.4}$ at mid-latitudes.   \citet{AugMathis2019} recently presented a model of rotating overshooting convection which employs the  semi-analytical model of \citet{Zahn91} and estimated the scaling of the overshoot depth  with the Rossby number. They also found that  $\delta$ decreases with decreasing Rossby number like $\delta\propto$ Ro$^{0.3}$. 

Overshooting in the Sun is thought to play a role in many aspects of the global flows and magnetic fields.  For instance, temperature variations arising from rotating convective overshoot are thought to be responsible for inducing a thermal wind in the convection zone that leads to the observed tilt of differential rotation contours \citep{RK95,Rempel2005,Miesch2006,Howe2009}.  Moreover, overshooting may pump magnetic field into the solar tachocline region where it can be stretched into large-scale toroidal field.   As a result, the tachocline has been the presumed seat of the solar dynamo for some time, figuring particularly prominently in the flux-transport and interface class of dynamo models \citep[e.g.,][]{Parker93,Dikpati1999,Rempel2006}.

For these reasons, a region of overshoot has been included in several 3D models of stellar convection.  This includes studies within the solar context \citep[e.g.,][]{Ghizaru2010,Racine2011,BMT11,Guerrero2013,Guerrero2016, Augustson2015}, as well as others related to more massive stars \citep[e.g.,][]{Browning2004,BBT2005,Featherstone2009,Aug12,Aug2016}.  Throughout those studies, however, the region of overshoot was but one ingredient in numerical experiments targeting other aspects of the convection and its resulting dynamo. For instance, \citet{Brun2017} studied solar-like overshooting convection but  mostly focused  on the mean flows arising within different rotational regimes  and the dependence of overshooting on stellar mass \citep[also see][]{BMT11,Aug12}. A systematic study of the effects of rotation on the convective overshooting dynamics, however, has yet to be performed in 3D global numerical calculations.

\par In this work, we use fully  non-linear 3D models of stratified convection in a spherical shell geometry to examine the dependence of convective overshooting on strong density stratification. We also develop scaling laws that describe the overshoot lengthscale with  Rossby number and investigate the properties of angular momentum transport and meridional flow in the coupled CZ-RZ system.  The paper is organized as follows. In Section \ref{sec:model}, we describe
our two-layered model formulation and provide the set of anelastic Navier-Stokes equations along with the initial conditions, and
the boundary conditions used in our simulations. In Section \ref{sec:Nrho}, we present our results associated with the effect of the density stratification on the non-rotating overshooting dynamics. In Section \ref{sec:rot}, we study  the dynamics related to the effect of rotation on the turbulent convective overshooting motions, as well as the dependence of the  overshoot lengthscale on the Rossby number and on latitude. In Section \ref{sec:am}, we focus on the rotational profiles arising at different Rossby numbers, the associated mean flows and the angular momentum transport and balances within the convection zone and the overshoot region. Finally, in Section \ref{sec:disc}, we summarize our results,   provide comparisons with previous  work, and discuss the implications of these results in the solar context.
\section{MODEL FORMULATION}
\label{sec:model}
\subsection{Dimensional Equations}
We aim to explore the dynamics associated with a two-zone system that consists of a convectively unstable region  that lies above a stably stratified radiative zone. For all cases presented in this work, we use a fixed aspect ratio  $r_i/r_o=0.45$ where $r_i$ is the inner radius  and $r_o$ is the outer radius of the spherical shell. Also, we assume that the convective region has a fixed depth $L=r_o-r_{c}=0.2408r_o$, where $r_{c}=0.7592r_o$ is the radius at the bottom of the CZ. Our chosen depth $L$ corresponds to the  solar convection zone from $\sim 0.7187R_{\odot}$ to $\sim 0.9467R_{\odot}$, hence only accounting for its  inner part and not the outer layers where radiative transfer processes take place.
We are interested in the dynamics within deep stellar interiors where the flows are subsonic. As such, we employ the anelastic approximation which filters out the sound waves and assumes small perturbations of the thermodynamic variables compared with their mean {\color{black} \citep{anelasticG,anelasticGG}}. Therefore, we solve the 3D Navier-Stokes equations {\color{black} under the anelastic and Lantz-Braginsky-Roberts approximations \citep{anelastic1,anelastic2}, the latter of which is exact in the convection zone where the reference state is adiabatic.}

Then, the dimensional Navier-Stokes equations become
\begin{equation}
\label{eq:momeq}
\dfrac{\partial{\vel}}{\partial{t}} +\vel\cdot\nabla\vel+2\Omega_o \hat{z}\times\vel=\dfrac{g(r)}{c_p}S\boldsymbol{\hat{r}}-\nabla(P/\bar{\rho})+\dfrac{1}{\bar{\rho}}\nabla\cdot \bold{D},
\end{equation}
\begin{equation}
\label{eq:conteq}
\nabla\cdot(\bar{\rho}\vel)=0,
\end{equation}

\begin{equation}
\label{eq:energyeq}
\bar{\rho}\bar{T}\left(\dfrac{\partial{S}}{\partial{t}}+\vel\cdot\nabla S +u_r\dfrac{d\bar{S}}{dr}\right)=\nabla\cdot(\bar{\rho}\bar{T}\kappa\nabla S)+Q+\bold{\Phi},
\end{equation}
where $\vel=(u_r,u_{\theta},u_{\phi})$ is the velocity field,  $P$ is the pressure, $\bar{\rho}$ is the reference density,  $\bar{T}$ is  the reference  temperature,  $d\bar{S}/dr$ is the background entropy gradient, $S$ characterizes the entropy perturbations about the reference state, $\Omega_o$ is the frame rotation rate, $g(r)$ is the gravity (where $g\propto 1/r^2$), and $c_p$ is the specific heat at constant pressure. The viscosity $\nu$ and the thermal diffusivity $\kappa$  are kept constant for simplicity. The viscous stress tensor $\bold{D}$ is given by
\begin{equation}
\bold{D}=2\bar{\rho}\nu(e_{ij}-\dfrac{1}{3}\nabla\cdot\vel),
\end{equation}
and the viscous heating is denoted by 
\begin{equation}
\bold{\Phi}=2\bar{\rho}\nu[e_{ij}e_{ij}-\dfrac{1}{3}(\nabla\cdot\vel)^2],
\end{equation}   
where $e_{ij}$ is the strain rate tensor. The internal heating term $Q$  is proportional to the pressure  such that $L_{\odot}=4\pi\int_{r_i}^{r_o} Q(r)r^2 dr$, where $L_{\odot}$ is the solar luminosity.  As noted in \cite{Featherstone16a}, this   profile is similar in structure to the one established in Model S \citep[see e.g.,][]{ModelS}. We taper the profile using a tanh function below the base of the CZ such that
\begin{equation}
\label{eq:Q}
Q(r)\propto \dfrac{1}{2}\left(\tanh\left(\dfrac{r-r_{c}}{0.01r_{c}}\right)+1\right)(\bar{P}(r)-\bar{P}(r_o)),
\end{equation}
where $\bar{P}$ is the reference pressure. It satisfies $Q(r)=-\nabla\cdot F_{rad}$, where $F_{rad}$ is   the radiative flux in the system  given by
\begin{equation}
\label{eq:Frad}
F_{rad}(r)=\dfrac{1}{r^2}\int_{r}^{r_o}Q(r')r'^2dr'.
\end{equation}
To close the set of equations (\ref{eq:momeq})-(\ref{eq:energyeq}), we use an equation of state  given by
\begin{equation}
\label{eqstate}
\dfrac{\rho}{\bar{\rho}}=\dfrac{P}{\bar{P}}-\dfrac{T}{\bar{T}}=\dfrac{P}{\gamma\bar{P}}-\dfrac{S}{c_p},
\end{equation}
assuming the ideal gas law
\begin{equation}
\label{eq:gaslaw}
\bar{P}=\Re\bar{\rho}\bar{T},
\end{equation}
where $\rho$  ($T$) are the density (temperature) perturbations about the reference state, $\Re$ is the gas constant, and $\gamma=c_p/c_v$ is the heat capacity ratio (where $c_v$ is the specific heat at constant volume).

\par To create our two-layered system, we  can then select an appropriate $d\bar{S}/dr$ profile that satisfies an adiabatic polytropic solution with $\gamma=5/3$  in the convective region $r_{c}\leq r\leq r_o$ and which smoothly transitions to  a stably stratified region for $r_i\leq r< r_{c}$. {\color{black}  We note however, that there are more sophisticated approaches where the two-zone system can be created by assuming that the radiative heat flux depends on a varying (in density and temperature) thermal conductivity profile based on either computed and tabulated opacities or on approximations such as Kramers law  \citep[see e.g.,][]{Kapyla2019,Kapyla2019GAFD,Viviani21}.} 

We also specify the number of density scale-heights in the convection zone through  $N_{\rho}=\ln(\bar{\rho}(r_c))/\bar{\rho}(r_o))$ (see Appendix \ref{sec:appendixA}). 
Then, we integrate $d\bar{S}/dr$ from $r_c$ inward, yielding a solution for  $\bar{S}$. The integration constant is computed  by matching the solution to the value of $\bar{S}$ at   the base of the CZ.  \\
Using the fact that the entropy for a monatomic ideal gas is   $\bar{S}=\ln(\bar{P}^{1/\gamma}/\bar{\rho})$,  differentiating the latter with respect to the radius, and  applying the hydrostatic balance equation ${\partial \bar{P}}/{\partial r}=-\bar{\rho} {g}$ we obtain
\begin{equation}
\label{dsdrprofile}
\dfrac{\bar{\rho}}{c_p}\dfrac{{d\bar{S}}}{dr}+\dfrac{{g}}{\gamma}\exp({{-\gamma \bar{S}}/{c_p}})\bar{\rho}^{(2-\gamma)}+\dfrac{d\bar{\rho}}{dr}=0,
\end{equation}
which we can  solve numerically for the reference density profile $\bar{\rho}(r)$ given any background entropy gradient profile. 

\subsection{Non-dimensional Equations}

We non-dimensionalize equations (\ref{eq:momeq})-(\ref{eq:energyeq}) using the depth of the convection zone $[l]=L$ as the lengthscale, the viscous timescale $[t]=L^2/\nu$, and the velocity scale $[u]=\nu/L$. Throughout the paper, we define the volume average of a quantity $q(r,\theta,\phi)$ in the CZ as 
\begin{equation}
\label{eq:volCZ}
{\color{black}{{q}_{cz}=\dfrac{\int_{r_{c}}^{r_o}\int_0^{2\pi}\int_0^{\pi} q r^2\sin\theta d\theta d\phi dr}{\int_{r_{c}}^{r_o}\int_0^{2\pi}\int_0^{\pi}  r^2\sin\theta d\theta d\phi dr}}.}
\end{equation} For the density and temperature scales, we choose
$[\rho]=\bar{\rho}_{cz}$, and $[T]=\bar{T}_{cz}$, respectively.

For  the entropy perturbations $S$, we adopt a scaling related to the thermal energy flux $F$ such that $[S]={L{F}_{cz}}/({{\bar{\rho}}_{cz}{\bar{T}}_{cz}\kappa})$, where $F(r)=\int_{r_i}^{r}Q(r')r'^2dr'/{r^2}$.
We  specify our chosen non-dimensional $d\bar{S}/dr$ profile in the RZ  such that
\begin{equation}
\label{eq:dsdr}
\dfrac{d\bar{S}}{dr}=\dfrac{A}{2}\left(1-\tanh\left(\dfrac{r-r_{b}}{d}\right)\right),
\end{equation}
where $A=10$, $r_{b}=0.6453r_o$ and $d=0.0456r_o$.
 
The non-dimensional gravity (reference  density $\bar{\rho}$, reference temperature $\bar{T}$) is equivalent to the dimensional gravity (reference density, reference temperature) divided by $g_{cz}$ ($\bar{\rho}_{cz}$, $\bar{T}_{cz}$). All the quantities are now non-dimensional  and the  non-dimensional anelastic Navier-Stokes equations become
\begin{equation}
\label{eq:NDmom}
\dfrac{\partial{\vel}}{\partial{t}} +\vel\cdot\nabla\vel+\dfrac{2}{\rm{Ek}}\hat{z}\times\vel=\dfrac{\rm{Ra}}{\rm{Pr}}g(r)S\boldsymbol{\hat{r}}-\nabla(P/\bar{\rho})+\dfrac{1}{\bar{\rho}}\nabla\cdot \bold{D},
\end{equation}

\begin{equation}
\label{eq:NDconteq}
\nabla\cdot(\bar{\rho}\vel)=0,
\end{equation}

\begin{equation}
\label{eq:NDenergyeq}
\bar{\rho}\bar{T}\left(\dfrac{\partial{S}}{\partial{t}}+\vel\cdot\nabla S +u_r\dfrac{d\bar{S}}{dr}\right)=\dfrac{1}{\rm{Pr}}\nabla\cdot(\bar{\rho}\bar{T}\nabla S)+\dfrac{1}{\rm{Pr}}Q_{nd}+\dfrac{\rm{Di Pr}}{\rm{Ra}}\bold{\Phi},
\end{equation}
where $Q_{nd}=LQ/{{F}_{cz}}$ {\color{black} is  the non-dimensional internal heating function. In Fig. \ref{fig:Q}, we show the profile of $Q_{nd}(r)$ against $r/r_o$ for the run with $N_{\rho}=3$.}

\Qfig

\noindent We now have four non-dimensional numbers: the flux Rayleigh number Ra, the Prandtl number Pr,  the Ekman number Ek and the  dissipation number Di, which are defined respectively as
\begin{equation}
\label{eq:RaPrEDi}
{\rm{Ra}}=\dfrac{{g}_{cz}{F}_{cz}L^4}{c_p\bar{\rho}_{cz}\bar{T}_{cz}\kappa^2\nu},\quad
{\rm{Pr}}=\dfrac{\nu}{\kappa},\quad
{\rm{Ek}}=\dfrac{\nu}{\Omega_oL^2},\quad\text{and}\quad 
{\rm{Di}}=\dfrac{{g}_{cz}L}{c_p\bar{T}_{cz}}.
\end{equation}
Note that, unlike Ra, Pr and Ek,  Di is not a free parameter. It depends on the chosen $N_{\rho}$ and  accounts for the fact that we have two energy scales -- a thermal scale and a kinetic scale. Thus, Di appears in the thermal equation to account for the viscous heating term, which is kinetic in nature.

We have run a series of 3D numerical simulations solving equations (\ref{eq:NDmom})-(\ref{eq:NDenergyeq}) using the Rayleigh  code \citep{Featherstone16a, Matsui16,Rayleighcode} with a chosen non-dimensional background entropy gradient $d\bar{S}/dr$ as described above and shown in Figure \ref{fig:dSdr}. 
\dsdrfig
We use impenetrable and stress-free boundary conditions for the velocity. For the entropy perturbations, we assume $\partial S/\partial r|_{r_i}=0$ at the inner boundary to account for the  fixed flux of energy generated due to the nuclear burning at the stellar core while we fix the entropy at the outer boundary such that $S|_{r_o}=0$. We note, however, that recent studies seem to suggest that  a fixed flux outer boundary condition might also be a  sensible choice in the stellar context \citep{Matilsky20}. In all simulations, we assume a fixed Pr $=1$ in order to avoid the onset of non-physical modes known to arise in rapidly rotating anelastic convective systems for Pr $< 1$ \citep[e.g.,][]{Calkins15}, while we vary $N_{\rho}$, Ra and Ek (see Table \ref{tab:table}).
Each simulation is evolved  from a zero initial
velocity and small-amplitude perturbations in the entropy field
until a statistically stationary and thermally relaxed state is reached.

\section{Effect of   density stratification  on the non-rotating overshooting dynamics}
\label{sec:Nrho}
We begin by exploring the effect of the density stratification in the CZ quantified by $N_{\rho}$ on the overshooting dynamics in a non-rotating spherical shell.
In Fig. \ref{fig:uNrho}, we present snapshots of meridional slices of the radial velocity $u_r$ at Ra $=10^5$ for increasing values of  $N_{\rho}$, ranging from $N_{\rho}=0.01$   to $N_{\rho}=4$. From a  visual inspection, we  see that,  in all of these cases, the convective motions generated in the unstable CZ do not  stop at the bottom of the CZ (marked by the inner black line), but they  travel some distance into the stable RZ which does  seem to depend on $N_{\rho}$, at least qualitatively. Furthermore, we notice that for increasing values of $N_{\rho}$,  the velocity field becomes more asymmetrical with narrower lanes of downflows and wider lanes of upflows. This enhanced asymmetry  is  a result of the corresponding increased density stratification in the CZ. By contrast, as  $N_{\rho}\rightarrow 0$, we are approaching the Boussinesq approximation limit, where the density is almost constant throughout the shell. Thus, the asymmetry for lower values of $N_{\rho}$ is mostly a result of  the sphericity and the mixed thermal boundary conditions \citep[e.g.,][]{Korre17}, though \cite{Anders2020} recently showed that the latter do not  substantially break the symmetry in the bulk of the convective region.
\ursNrhofig
\par There are many different ways of examining quantitatively how far these convective motions can overshoot into the stable region.  In this work we will rely on the radial variation of the  kinetic energy to estimate overshooting lengthscales.
Throughout the paper, we define the time and spherical average of a quantity $q$ as 
\begin{equation}
\label{eq:qav}
\tilde{q}(r)=\dfrac{1}{4\pi(t_2-t_1)}\int_{t_1}^{t_2}\int_0^{2\pi}\int_0^{\pi}q(r,\theta,\phi,t)\sin\theta d\theta d\phi dt,
\end{equation}
{\color{black}
and the time and azimuthal average of a quantity $q$ as
\begin{equation}
\label{eq:qaz}
\langle q(r,\theta) \rangle=\dfrac{1}{2\pi(t_2-t_1)}\int_{t_1}^{t_2}\int_0^{2\pi}q(r,\theta,\phi)d\phi dt,
\end{equation}
}
where $t_2-t_1$ is a time interval in the simulation in a statistically stationary and thermally equilibrated state. 
\par In Figure \ref{fig:KE_Nrho_allfig}a, we plot the time- and spherically- averaged kinetic energy  $\tilde{E}_k(r)$ (where {\color{black}{$\tilde{E}_k(r)=(1/2)\bar{\rho}(\widetilde{u_r^2}+\widetilde{u_{\theta}^2}+\widetilde{u_{\phi}^2}$}})) for different $N_{\rho}$ values at Ra $=10^5$. We verify what we noticed in Fig. \ref{fig:uNrho}, namely that $\tilde{E}_k(r)$ is large in the CZ, but there is also  substantial kinetic energy below the base of the convective region, where the convective motions overshoot. However, these motions do decelerate within the stable RZ, where they are no longer buoyantly driven there, and $\tilde{E}_k$ decreases accordingly.
We notice that  $\tilde{E}_k$  acquires a half-Gaussian profile below $r_{c}$, similarly to what \cite{Korre19} recently   found in their Boussinesq convective overshooting spherical shell simulations. Indeed, this seems to be a salient characteristic of the overshooting dynamics, independent of the approximation (here we assume the anelastic approximation) and  the model set-up employed. Using this feature,  we fit a Gaussian function of the form
\begin{equation}
\label{eq:fG}
f_G(r)=A_E\exp\left(-\left(\dfrac{r-r_{c}}{\sqrt{2}\delta_{Gf}}\right)^2\right)
\end{equation}
to the $\tilde{E}_k(r)$ data {\color{black}from the bottom of the CZ $r_{c}$ down to the point where $\tilde{E}_k(r)$ is well-fit by Eq. (\ref{eq:fG}) and use the computed width of this Gaussian $\delta_{G}=\delta_{Gf}/r_o$}  as an estimate of   how far the turbulent convective motions overshoot in the stable region, in an average sense \citep[see for details][]{Korre19}. In Figure \ref{fig:KE_Nrho_allfig}b, we plot $\tilde{E}_k(r)$ for a typical run (with $N_{\rho}=3$)  along with the fitted  function $f_G(r)$ given in Eq. (\ref{eq:fG}) to illustrate how well a Gaussian of this form actually fits the numerical data with $\delta_G=0.049$ for this case.
\KE_Nrho_allfig
 
\par In order to examine the dependence of the overshooting lengthscale on the density stratification, in Fig. \ref{fig:deltaGNrho}, we plot $\delta_G$ against $N_{\rho}$ and  notice that  $\delta_G$ increases with increasing $N_{\rho}$ for $N_{\rho}\lesssim 0.7$, while it  starts decreasing with increasing $N_{\rho}$ for $N_{\rho}>0.7$.   This result might initially seem counter-intuitive, since we would expect that the enhanced asymmetry, which leads to  narrower downflows for  larger values of $N_{\rho}$ (as also seen in Fig. \ref{fig:uNrho}), would  result in larger values of  $\delta_G$ \citep[see for instance][]{Mass84}. However,  it is crucial to also consider the density stratification in the RZ which is not the same for the different $N_{\rho}$ cases in our simulations.
 We found  that the relevant quantity is the ratio of the density stratification in the RZ over the one in the CZ, given by 
 {\color{black}
 \begin{equation}
 \label{eq:Rn}
 R_{\rho}=\dfrac{(\bar{\rho}(r_i)/\bar{\rho}(r_c))}{(\bar{\rho}(r_c)/\bar{\rho}(r_o))},
 \end{equation}
 }
and not solely $N_{\rho}$.
In Figure \ref{fig:dgRn}a, we plot the computed $\delta_G$ against $R_{\rho}$ and find that $\delta_G$  decreases with  decreasing $R_{\rho}$ such that $\delta_G\propto R_{\rho}^{0.36}$. This indicates that $R_{\rho}$ is a more sensible parameter for  explaining our results. Indeed, the density ratio in the CZ is explicitly associated with both the degree of asymmetry of the flow and the strength of the convective motions with increased stratification leading to somewhat smaller kinetic energies in the CZ\footnote{Note that the  magnitude of the kinetic energy decreases somewhat with increasing $N_{\rho}$ due to the fact that all cases shown in Section \ref{sec:Nrho} use Ra $=10^5$, but with increasing $N_{\rho}$, the critical Rayleigh number also increases \citep[see e.g.,][]{Jones2009, Gastine2012}, so they do not effectively have the exact same supercriticality.}.
On the other hand, the density ratio in the RZ dictates how strongly stably stratified that region is. We thus conclude that the overshoot lengthscale  will depend on the density ratios in both the CZ and the RZ which can be described by $R_{\rho}$. 

\par Finally, in Figure \ref{fig:dgRn}b, we present a snapshot of the radial velocity for the case of $N_{\rho}=3$ and Ra $=10^5$ at a chosen longitude against $r/r_o$ and $\theta$ along with the distance down to $r_c-\delta_G$ marked with a blue solid line  to illustrate the computed overshoot distance from the bottom of the CZ (blacked dashed line). We find that $\delta_G$  marks the distance down to which the convective motions overshoot in an average sense.
\deltaGNrhofig
\dgRnfig
We now briefly focus on the ``penetrative" dynamics associated with thermal mixing in the stable region. To do so, we define the  time- and spherically- averaged total adjusted entropy gradient profile $d\tilde{S}_T/dr$ such that
\begin{equation}
\label{eq:dSNrho}
\dfrac{d\tilde{S}_T}{dr}=\dfrac{d\bar{S}}{dr}+\dfrac{d\tilde{S}}{dr},
\end{equation}
i.e. the sum of the reference entropy gradient and the gradient of the entropy fluctuations.
In Figure \ref{fig:dSNrho}, we plot the background reference  $d\bar{S}/dr$ profile and compare it to the adjusted  $d\tilde{S}_T/dr$ for four $N_{\rho}$ cases spanning a wide range of  density stratifications.
We find that there is some partial thermal mixing in the RZ.  Indeed, $d\tilde{S}_T/dr$ deviates somewhat from the background $d\bar{S}/dr$ there and exhibits some slight dependence on $N_{\rho}$. However,   $d\tilde{S}_T/dr>0$ in all cases, namely the CZ does not further extend into the stable zone, so no pure penetration is observed. In fact, there appears to be a slightly subadiabatic layer near the base of the convective region. Although this is not a standard feature  in 3D overshooting simulations, it has been previously reported in the 3D compressible overshoot simulations of \citet{Kapyla} for the cases where  they employed a heat conduction
profile,  based either on a Kramers-like opacity or a static profile of similar shape, which varied smoothly between the CZ and the RZ. Thus, we attribute the subadiabatic layer observed in our runs to our chosen background entropy gradient profile  which transitions  smoothly from the CZ down to the RZ as well as to the fact that at this Ra, the degree of buoyant driving is not sufficiently strong  to establish a fully adiabatic convective interior.

\dSNrhofig

\begin{deluxetable}{cccccccccccccc}
\label{tab:table}
\centering
\tablecolumns{10}
\tablewidth{0pc}
{\color{black}

\tablecaption{Input and output parameters of the simulations}
\tablehead{
\colhead{Case} & \colhead{$N_{\rho}$} & \colhead{Ra} & 
& \colhead{Ek}  & \colhead{Ra$_{r}$} &  \colhead{Ro$_c$} & $N_r\times N_{\theta}\times N_{\phi}$ & \colhead{Ro} & \colhead{$\delta_G$} & \colhead{Re}}
\startdata
1 & 0.01 & $10^5$ & 
& \nodata & \nodata & \nodata & 192$\times$528$\times$1056 & \nodata & 0.081  & 44.05 \\
2 & 0.1 & $10^5$ & 
& \nodata   & \nodata & \nodata & 192$\times$528$\times$1056  & \nodata & 0.083 & 43.55 \\
3 & 0.3 & $10^5$ & 
& \nodata  &  \nodata & \nodata & 192$\times$528$\times$1056 & \nodata  & 0.090 & 42.44 \\
4 & 0.5 & $10^5$ & 
& \nodata   &  \nodata & \nodata & 192$\times$528$\times$1056 & \nodata & 0.091 & 41.58 \\
5 & 0.7 & $10^5$ & 
& \nodata   & \nodata & \nodata & 192$\times$528$\times$1056 & \nodata & 0.091 & 40.37  \\
6 & 0.9 & $10^5$ & 
& \nodata   &  \nodata &  \nodata & 192$\times$528$\times$1056  & \nodata &  0.088 & 39.48 \\
7 & 1 & $10^5$ & 
& \nodata  & \nodata & \nodata & 192$\times$528$\times$1056 & \nodata & 0.087 & 39.08 \\
8 & 2 & $10^5$ & 
&   \nodata   & \nodata & \nodata & 192$\times$528$\times$1056 & \nodata  & 0.064 & 35.60 \\
8a & 2 & $10^5$ & 
&   0.001 & 10 & 0.158 & 192$\times$528$\times$1056  & 0.0072 &  0.014 & 14.33 \\
8b & 2 & $10^5$ & 
&   0.01 & 215.4 & 1.58 & 192$\times$528$\times$1056  & 0.2627 & 0.045 & 52.53 \\
8c & 2 & $10^5$ & 
&   0.1  & 4642 & 15.8 & 192$\times$528$\times$1056 & 1.8066 & 0.061 & 36.13 \\
9 & 3 & $10^4$ & 
&   \nodata  & \nodata & \nodata & 192$\times$256$\times$512 & \nodata  & 0.049 & 13.31\\
9a & 3 & $10^4$ & 
&   0.01 & 21.5 &  0.5 & 192$\times$256$\times$512 & 0.0423  & 0.018  & 8.46 \\
9b & 3 & $10^4$ & 
&   0.1  & 464.2 & 5 &192$\times$256$\times$512 & 0.6855 & 0.048 &  13.71  \\
10 & 3 & $10^5$ & 
&   \nodata  & \nodata & \nodata & 192$\times$528$\times$1056  & \nodata & 0.049 & 33.54 \\
10a & 3 & $10^5$ & 
&   0.001  & 10 & 0.158 &192$\times$792$\times$1584 & 0.0067 & 0.013 & 13.47 \\
10b & 3 & $10^5$ & 
&   0.01  & 215.4 & 1.58 & 192$\times$528$\times$1056 & 0.2263 &  0.039 & 45.26  \\
10c & 3 & $10^5$ & 
&   0.1  & 4642 & 15.8 & 192$\times$528$\times$1056 & 1.6991 &  0.049  & 33.98 \\
11 & 3 & $10^6$ & 
&   \nodata & \nodata & \nodata & 192$\times$1104$\times$2208 & \nodata  & 0.048 & 81.20\\
11a & 3 & $10^6$ & 
&   0.001  & 100 & 0.5 & 192$\times$1104$\times$2208 & 0.0432 & 0.028 & 86.39 \\
11b & 3 & $10^6$ & 
&   0.01  & 2154 & 5 & 192$\times$1104$\times$2208 &  0.5093 & 0.047 & 101.86 \\
12 & 4 & $10^5$ & 
&   \nodata  & \nodata & \nodata & 192$\times$528$\times$1056 & \nodata & 0.042 & 31.87 \\
\enddata

\tablecomments{{\color{black} Columns $2-6$ indicate the input  parameters, column $7$ provides the resolution  and columns $8-10$ report on the output parameters for both the non-rotating and the rotating simulations.}}}
\end{deluxetable}

\section{Effect of rotation on the overshooting dynamics}
\label{sec:rot}
\par Our work is motivated by  {\color{black}solar-type} stars, and so we are particularly  interested in studying the effect of rotation on the overshooting dynamics. The solar tachocline, for instance, coincides with the region of overshoot near the base of the solar convection zone, and so understanding the effects of rotation within that region represents an important step toward gaining a better understanding of the tachocline dynamics as well. In this Section, we investigate the effect of rotation on the convective overshooting dynamics by considering a series of numerical experiments where we vary $N_{\rho}$, Ra and Ek. To categorize each case and the corresponding rotational regime, we output the Rossby number Ro defined as
\begin{equation}
\label{eq:Ronum}
{\rm{Ro}}=\dfrac{U}{2 L \Omega_0}=\dfrac{{\rm{Re}}{\rm{Ek}}}{2},
\end{equation} 
 which is the ratio of inertial forces to the Coriolis force. Here, the first expression includes dimensional quantities where  $U$ is a typical dimensional velocity in the CZ, while the second expression includes non-dimensional quantities where $\rm{Re}=UL/\nu$ is the Reynolds number  extracted from the simulations (where $U=u_{rms}\nu/L$). For the value of $u_{rms}$, we adopt the total (fluctuating part and mean part)  {\color{black} velocity averaged over the CZ and weighted by density}. Note that the P\'{eclet} number Pe is  Pe=PrRe, so it is simply Pe=Re in our Pr $=1$ runs. \\
{\color{black} In Table \ref{tab:table}, we also report on the reduced Rayleigh number Ra$_r=$RaEk$^{4/3}$ which is a typical measure of supercriticality in simulations of rotating convection as well as on the convective Rossby number Ro$_c=\sqrt{\rm Ra \rm Ek^2 /(4\rm Pr)}$ which is a measure of the influence of rotation on the global flows  based on the input parameters Ra, Pr and Ek.}
\par We begin by qualitatively examining the flow characteristics for models with four representative values of the resulting Ro. In Fig. \ref{fig:ur_shell_rot}, we present snapshots of shell slices of the radial velocity $u_r$ close to the outer boundary of the spherical shell at $r=0.9r_o$, for cases with Ro spanning values from $0.007$ up to $1.7$. We notice that for the  highest values of Ro, the convective flow is not significantly affected by rotation with the Ro $\approx 0.23$ case exhibiting only  a slight alignment of the flow with the axis of rotation within the equatorial region. For the smaller Ro $\approx 0.043$ run, the flow is more columnar near the equator and  mid-latitudes while the plumes  observed closer to the poles indicate that inertia is dominant there. For the lowest Ro case of Ro $\approx 0.007$, rotation dominates the flow and leads to a purely columnar profile at mid-latitudes and the equator, while $u_r$ is negligible at the poles.
  
To obtain a qualitative view of the fluid motions across the  shell,  we show snapshots of meridional slices of $u_r$ versus $\theta$ and $r$ at a fixed longitude for the same four computed Rossby numbers in Fig. \ref{fig:Rour}. We observe that for the highest Ro $\approx 1.7$ case, where the Coriolis force is relatively weak, convective motions overshoot a substantial distance below the base of the CZ. As Ro decreases to Ro $\approx 0.23$, the convective motions begin to align with the axis of rotation but still overshoot into the RZ. As Ro decreases further, convective motions increasingly align with the rotation axis, both in the convection zone and in the region of overshoot. Moreover, from a visual inspection, we find that, at least qualitatively, the convective motions seem to overshoot  smaller distances into the RZ with decreasing Ro. 
\ur_shell_rotfig

\Rourfig

\par We next compute the overshoot lengthscale $\delta_G$ for each run, following the  same process we used in Section \ref{sec:Nrho}. However, for these rotating cases, we fit the Gaussian function (Eq. (\ref{eq:fG})) to the fluctuating kinetic energy profile $\tilde{E}_f(r)$ (where {\color{black}{$\tilde{E}_{f}(r)=({1}/{2})\bar{\rho}[{\widetilde{u_{f,r}^2}+\widetilde{u_{f,\theta}^2}+\widetilde{u_{f,\phi}^2}]}$}}). This definition filters out the mean, axisymmetric flows that arise in rotating convection. Note that $\vel=\vel_m+\vel_f$ where $\vel_m$ is the mean, axisymmetric part of the velocity corresponding to the larger-scale motions associated with rotation and and $\vel_f$ is the fluctuating part of velocity field associated with the convective motions.

In Fig. \ref{fig:Rodov}, we present a plot of the values of $\delta_G$ along with their errorbars (arising from uncertainty in the fit) against Ro. We note that the errorbars are quite significant in the rotationally-constrained cases of Ro $<<0.01$ where the kinetic energy profiles are not well-fit by the Gaussian below $r_c$.  There, the weak convective motions have already begun to exponentially decay within the bulk of the CZ. Thus, we verify the qualitative behavior illustrated in Fig. \ref{fig:Rour} and find that $\delta_G$ decreases with decreasing Ro as  $\delta_G\propto$ Ro$^{0.23}$.
 
\par  The fact that $\delta_G$ decreases with decreasing Ro is  somewhat expected, since the enhanced horizontal interaction associated with rotation can lead to faster braking of the convective overshooting motions within the regimes where the Coriolis force plays a more dominant role. There, the convective flow  aligns with the rotation axis, forcing the downflows to overshoot from an angle. Then, the intrinsically radial buoyancy braking  in the RZ leads to a more efficient halting of the overshooting motions  since it only has to counteract their radial component. This can be viewed qualitatively in Fig. \ref{fig:Rour},  which clearly shows that the radial flow is more aligned with the axis of rotation for lower values of Ro.
\Rofig
More quantitatively, in Figure \ref{fig:deltaGtheta}, we illustrate the dependence of the computed overshoot lengthscale $\delta_G$ on the latitude $\theta$ at the same four representative Ro cases with Ro $\approx 0.007$, Ro $\approx 0.043$, Ro $\approx 0.23$ and  Ro $\approx 1.7$.
We find that for Ro $\approx 1.7$, $\delta_G$ only weakly depends on latitude, since  the influence of rotation on convection is insignificant. For the  Ro $\approx 0.23$ case, the Coriolis force is somewhat stronger, leading to a shallower $\delta_G$ overall which is larger closer to the poles and becomes smaller at mid-latitudes  where rotation becomes more dominant.  For Ro $\approx 0.043$,  $\delta_G$ is  considerably larger within the polar regions while it decreases as we approach the mid-latitudes and the equator, commensurate with the fact that the overshooting motions penetrate at an angle there due to  the alignment of the convective motions with the axis of rotation (see Fig. \ref{fig:Rour}b). Finally, for the rotationally-constrained case of Ro $\approx 0.007$, we only show $\delta_G$ near the equator, since  the extracted $\delta_G$  does not give meaningful results away from the equatorial region where the convective motions are  much weaker. We see that $\delta_G$ in this case is substantially smaller compared with the other Ro cases.

\deltaGthetafig

\FRofig

To further examine the dependence of the overshooting dynamics on Ro,
in Fig. \ref{fig:FRo}, we plot the  time- and spherically- averaged non-dimensional radial fluxes (multiplied by the surface area $4\pi r^2$), namely the kinetic energy flux  $\tilde{F}_k$,    the enthalpy flux  $\tilde{F}_e$,  the conductive flux  $\tilde{F}_c$, and  the viscous flux  $\tilde{F}_v$. These are respectively given by
\begin{equation}
\label{eq:Fk}
\tilde{F}_k(r)=\dfrac{\rm{DiPr}}{2\rm{Ra}}\bar{\rho}\widetilde{(u_r\vel^2}),
\end{equation}
\begin{equation}
\label{eq:Fe}
\tilde{F}_e(r)=\bar{\rho}(\widetilde{u_r T}),
\end{equation}
\begin{equation}
\label{eq:Fc}
\tilde{F}_c(r)=\bar{\rho}\bar{T}\left(\widetilde{\dfrac{\partial {S}}{\partial r}}\right),
\end{equation}
\begin{equation}
\label{eq:Fv}
\tilde{F}_v(r)=-\dfrac{\rm{DiPr}}{\rm{Ra}}(\widetilde{\vel\cdot\boldsymbol{D}})_r,
\end{equation}
for the four typical Rossby cases ranging from Ro $\approx 0.007$ to Ro $\approx 1.7$.  
{\color{black} On the same figure, we also show the radiative flux as given in Eq. (\ref{eq:Frad}) as well as the total flux
\begin{equation}
\label{eq:Ftot}
\tilde{F}_{tot}=F_{rad}+\tilde{F}_k+\tilde{F}_e+\tilde{F}_c+\tilde{F}_{\nu}.
\end{equation}  
}

We find that, compared with the lower Ro cases, the runs with Ro $\approx 1.7$ and Ro $\approx 0.23$ attain a substantially larger kinetic energy flux $\tilde{F}_k$ within the CZ and the overshoot region due to the fact that convection is very weakly affected by rotation (as also verified in Fig. \ref{fig:Rour} and Fig. \ref{fig:Rodov}). By contrast, in the lower Rossby case of Ro $\approx 0.043$,  $\tilde{F}_k$ is quite small within the CZ  while it becomes negligible at Ro $\approx 0.007$  whereby rotation has  dominated the dynamics. 
Similarly, $\tilde{F}_e$  starts decreasing within the CZ but the enthalpy flux is still non-zero below the base of the CZ where the convective motions overshoot. Overall, the magnitude of $\tilde{F}_e$  decreases with decreasing Ro, once again confirming that  convection becomes less efficient at larger rotation rates.   
Finally, $\tilde{F}_c$  increases with decreasing Ro  illustrating  that heat is mainly transported  by conduction once rotation becomes stronger. In all four cases however, $\tilde{F}_v$ is very small, suggesting that viscosity does not play a dominant role in the dynamics.

\newpage
\section{Mean flows and angular momentum balances}
\label{sec:am}

\subsection{Rotational Profiles and Mean Flows}

\par In Section \ref{sec:rot}, we focused on the fluctuating component of the kinetic energy in order to understand the dependence of the convective overshooting motions on Ro.
We now concentrate on the differential rotational profiles and the associated dynamics within the different regimes arising at different  Ro cases. We  also investigate the mean flows  resulting from velocity correlations due to the presence of rotation through the Coriolis force. 
In Fig. \ref{fig:MCDF}, we present slices of the time- and azimuthally- averaged  meridional circulation streamlines with the underlying contours of the mass flux (left panels) and  differential rotation profiles ($\langle u_{\phi}\rangle/(r\sin\theta)$) (right panels) for the four typical Ro cases. We see that for Ro $\approx 0.007$, there are two large  cells in the meridional circulation within the CZ and the overshoot region and four smaller ones close to the outer CZ boundary. All of these cells are localized at mid-latitudes and the equator where the flow motions are the strongest (also see Fig. \ref{fig:Rour}a), similarly to the differential rotation. We notice that the equator rotates faster than the poles, but convection in this case is inhibited by rotation, so it is in a rotationally-constrained regime. At Ro $\approx 0.043$, the profile exhibits a  multi-cellular flow near the equator that propagates into the RZ particularly within the equatorial region. The differential rotation is again much stronger at the equator and decreases at mid-latitudes and the poles in the CZ while the stable region rotates almost uniformly. This rotational profile is qualitatively similar to the one observed in the  solar CZ, although the latter is conical and not cylindrical as the one in this case. We will refer to this case as the ``solar-like" case. At Ro $\approx 0.23$, the meridional circulation has two distinct large  cells within the CZ,   while the differential rotation in the CZ is now characterized by faster rotating poles and a slower equator, and is more or less uniform in the bulk of the RZ. Hence,  we will refer to this case as ``anti-solar" \citep[e.g.,][]{Gastine14}. Finally, in the largest  Ro $\approx 1.7$ case,  the Coriolis force is very weak, and convection is not significantly affected by rotation. The meridional circulation consists of multiple cells and the convection zone now  rotates uniformly at a slower rate than the RZ. Overall, in all of these cases (except for the rotationally-constrained case of Ro $\approx 0.007$), there is significant propagation of the meridional circulation into the stable layer, even beyond the overshoot region.  

\MCDFRofig
To  understand this behavior more quantitatively, in Fig. \ref{fig:KEmeanflows}, we plot the time- and spherically- averaged non-dimensional kinetic energy  related to the differential rotation $\tilde{E}_{DR}$ and the meridional circulation $\tilde{E}_{MC}$ as a function of radius where{\color{black}
\begin{equation}
\label{eq:EDF}
\tilde{E}_{DR}(r)=\dfrac{1}{2}\bar{\rho}\widetilde{u_{m,\phi}^2},
\end{equation}
\begin{equation}
\label{eq:EMF}
\tilde{E}_{MC}(r)=\dfrac{1}{2}\bar{\rho}(\widetilde{u_{m,r}^2}+\widetilde{u_{m,\theta}^2}).
\end{equation}}
\KEmeanflowsfig
In Fig. \ref{fig:KEmeanflows}a, we see that the kinetic energy associated with the differential rotation $\tilde{E}_{DR}(r)$ is largest for the solar-like case and the anti-solar case both in the CZ and the RZ. It becomes very small at  Ro $\approx 1.7$, where the rotational influence is almost negligible, and in the lowest Ro $\approx 0.007$ case where the motions are very weak overall and are unable to redistribute angular momentum as efficiently. Moreover, in all four cases, $\tilde{E}_{DR}$ decreases with depth in the CZ,  but there is still considerable kinetic energy  below the base of the CZ (with the exception of Ro $\approx 0.007$).

As shown in Fig. \ref{fig:KEmeanflows}b,  the kinetic energy associated with the meridional flows $\tilde{E}_{MC}(r)$  decreases with decreasing Ro  in the CZ  and is very small for the lowest Ro $\approx 0.007$ case  (both in the CZ and in the RZ) where rotation dominates the dynamics. For the higher Ro cases, there is substantial kinetic energy in the meridional flows within the overshoot region and part of  the stable zone. This indicates that these larger-scale mean flows, although mostly generated in the CZ, do not merely stop at the base of the convective region, but they travel deeper  into the radiative zone.

\subsection{Angular Momentum Transport}

We now focus on the  solar-like case and the anti-solar case with Ro $\approx 0.043$ and Ro $\approx 0.23$ respectively.
We look at the angular momentum transport and the main  {\color{black} processes} responsible for redistributing angular momentum within the two-zone spherical shell.  Following \citet{Elliott2000} and {\color{black}\citet{BMT2004}}, by taking the $\phi$ component of the momentum equation, and assuming a statistically stationary state, averaged in time and longitude, we obtain 
\begin{equation}
\label{eq:AM}
\dfrac{1}{r^2}\dfrac{\partial(r^2 \langle F_r(r,\theta)\rangle)}{\partial r}+\dfrac{1}{r\sin\theta}\dfrac{\partial(\sin\theta  \langle F_{\theta}(r,\theta)\rangle)}{\partial\theta}=0.
\end{equation}
{\color{black}
 The radial flux   $ F_r(r,\theta)$  and the latitudinal flux $F_{\theta}(r,\theta)$ can be expressed non-dimensionally as
\begin{equation}
\label{eq:FrAMND}
 F_r(r,\theta)=\bar{\rho}r \sin\theta\left({{u_{f,r} u_{f,\phi}}}+{{u}_{m,r}{u}_{m,\phi}}+{\dfrac{{u}_{m,r} r\sin\theta}{{\rm Ek}}}+r{\dfrac{\partial}{\partial r}\left(-\dfrac{{u}_{m,\phi}}{r}\right)}\right),
\end{equation}
and
\begin{equation}
\label{eq:FthetaAMND}
 F_{\theta}(r,\theta)=\bar{\rho}r \sin\theta\left({{u_{f,\theta} u_{f,\phi}}}+{{u}_{m,\theta}{u}_{m,\phi}}+{\dfrac{{u}_{m,\theta} r\sin\theta}{{\rm Ek}}}+{\dfrac{\sin\theta}{r}\dfrac{\partial}{\partial \theta}\left(-\dfrac{{u}_{m,\phi}}{\sin\theta}\right)}\right),
\end{equation}
respectively. We  can then consider the time- and azimuthal- average  of the  individual terms in Eq. (\ref{eq:FrAMND}) such that
\begin{eqnarray}
\label{eq:termsAMr1}
F_{r,R}=\langle\bar{\rho}r \sin\theta({u_{f,r} u_{f,\phi}})\rangle,\\
\label{eq:termsAMr2}
F_{r,M}=\langle\bar{\rho}r \sin\theta({u}_{m,r} {u}_{m,\phi})\rangle,\\
\label{eq:termsAMr3}
F_{r,C}=\left\langle\bar{\rho}r \sin\theta\left(\dfrac{{u}_{m,r} r\sin\theta}{{\rm Ek}}\right)\right\rangle,\\
\label{eq:termsAMr4}
F_{r,V}=\left\langle\bar{\rho}r \sin\theta\left(-r\dfrac{\partial}{\partial r}\left(\dfrac{{u}_{m,\phi}}{r}\right) \right)\right\rangle.
\end{eqnarray}
In a similar manner, the individual terms in Eq. (\ref{eq:FthetaAMND}) are 
\begin{eqnarray}
\label{eq:termsAMthetaov1}
F_{\theta,R}=\langle\bar{\rho}r \sin\theta({u_{f,\theta} u_{f,\phi}})\rangle,\\
\label{eq:termsAMthetaov2}
F_{\theta,M}=\langle\bar{\rho}r \sin\theta({u}_{m,\theta}{u}_{m,\phi})\rangle,\\
\label{eq:termsAMthetaov3}
F_{\theta,C}=\left\langle\bar{\rho}r \sin\theta\left(\dfrac{{u}_{m,\theta} r\sin\theta}{{\rm Ek}}\right)\right\rangle,\\
\label{eq:termsAMthetaov4}
F_{\theta,V}=\left\langle\bar{\rho}r \sin\theta\left(-\dfrac{\sin\theta}{r}\dfrac{\partial}{\partial \theta}\left(\dfrac{{u}_{m,\phi}}{\sin\theta}\right)\right)\right\rangle.
\end{eqnarray} 
}
Note that Equations (\ref{eq:termsAMr1}) and (\ref{eq:termsAMthetaov1})  are associated with the Reynolds stresses, Equations (\ref{eq:termsAMr2}) and (\ref{eq:termsAMthetaov2}) and Equations (\ref{eq:termsAMr3}) and (\ref{eq:termsAMthetaov3}) correspond to the mean flows associated with the meridional circulation and finally Equations (\ref{eq:termsAMr4}) and (\ref{eq:termsAMthetaov4}) are related to the viscous stresses.

In Fig. \ref{fig:AMsolar}a,b, c and d we show the profiles of the latitudinal fluxes  at Ro $\approx 0.043$ (solar-like case). We observe that $F_{\theta,C}$ is  large near the equator and at  mid-latitudes within the CZ. There is substantial penetration in the stable region where the distinct cells have opposite signs commensurate with the meridional circulation profile in Fig. \ref{fig:MCDF}c. Angular momentum in the convection zone and in the overshoot region seems to be dictated by the flux related to the action of the Coriolis force on the mean meridional flows while all the other fluxes are comparatively very small.  In Fig. \ref{fig:AMsolar}e,f,g and h, we notice that  the radial flux  $F_{r,C}$ is again  large within the CZ while it also extends further down into the RZ transporting angular momentum mainly inward. The profile of $F_{r,R}$ shows that transport by Reynolds stresses is  weak compared to the transport by $F_{r,C}$, however it does contribute to the overall angular momentum transport with inward transport at mid-latitudes and outward transport near the equator. The flux associated with the mean flows $F_{r,M}$ is negligible, similarly to $F_{\theta,M}$, while $F_{r,V}$ is also  small overall although it does seem to become stronger close to the equator within the CZ where it transports angular momentum inward. 
\par For the anti-solar case of Ro $\approx 0.23$, in Figure \ref{fig:AMantisolar}c, we see that, similarly to the solar-like case, $F_{\theta,C}$ is dominant within the CZ and the overshoot region, while $F_{\theta,R}$, $F_{\theta,M}$ and $F_{\theta,V}$ are all much weaker (Fig. \ref{fig:AMantisolar}a,b and d). In Figure \ref{fig:AMantisolar}e,f,g and h, we  notice that $F_{r,C}$ is again the largest in magnitude compared with the other fluxes carrying angular momentum outward near the equator and inward at mid-latitudes while $F_{r,R}$ although not as large, is nonetheless  present within the CZ carrying angular momentum inward (except at the poles where it is too weak), similarly to the much weaker $F_{r,M}$. 
Finally, $F_{r,V}$ is mostly larger close to the base of the CZ within the equatorial region transporting angular momentum in the same direction as $F_{r,C}$. 
\AMsolarfig
\AMantisolarfig

\subsection{Thermal-wind Balance  within the CZ and the Overshoot Region}
{\color{black} We are interested in understanding what balances are achieved and how the mean flows are sustained  in a steady state across the two-zone spherical shell.} 
Two dominant terms that contribute to the meridional flows are the  rotational shear associated with the differential rotation and the baroclinicity of the mean stratification. This balance, known as thermal-wind balance, occurs when the system is in both hydrostatic and geostrophic equilibrium, namely when  there is a balance between the Coriolis force, the pressure and the gravity {\color{black}\citep[see e.g.,][]{R89,RK95,Rempel2005, Miesch2006,MH2011, AG2013}}.
Within the anelastic approximation, the dimensional thermal-wind equation is then given by
\begin{equation}
\label{eq:TWB1}
\dfrac{\partial \Omega^2}{\partial z}\approx\dfrac{g}{\lambda r c_p}\dfrac{\partial\langle S \rangle}{\partial \theta},
\end{equation}
where $z=r\cos\theta$, $\lambda=r\sin\theta$ and $\Omega=\Omega_o+\langle u_{\phi}\rangle/\lambda$ is the total angular velocity.
Non-dimensionally, Eq. (\ref{eq:TWB1}) can be expressed as
\begin{equation}
\label{eq:TWB2}
\underbrace{\dfrac{2}{\rm Ek}\left(r\cos\theta\dfrac{\partial\langle u_{\phi} \rangle}{\partial r}-\sin\theta\dfrac{\partial \langle u_{\phi} \rangle}{\partial \theta}\right)}_{{\rm{LHS}}}\approx \underbrace{\dfrac{{\rm Ra} g(r)}{\rm Pr}\dfrac{\partial \langle S \rangle}{\partial \theta}}_{{\rm{RHS}}},
\end{equation}
(where again $\langle\cdot\rangle$ denotes a time- and azimuthally- averaged quantity).
In Fig. \ref{fig:TWB}, we examine the thermal-wind balance for the solar-like case with Ro $\approx 0.043$ and the anti-solar case with Ro $\approx 0.23$. In Fig. \ref{fig:TWB}a, we plot the left-hand side term (LHS) term and right-hand-side (RHS) term of Eq. (\ref{eq:TWB2}) at different radii from the bottom of the CZ $r_c$ down to  $\sim r_c-\delta_G$ for the solar-like case. The LHS term is associated with the Coriolis force and thus refers to the inertia of the differential rotation,  while the RHS term  accounts for the baroclinicity of the stratification through the latitudinal gradient of the entropy perturbations. The thermal-wind balance is satisfied within the overshoot region for this solar-like case where LHS$\approx$RHS in that region. In Figure \ref{fig:TWB}b, we also show contours of the  LHS and  RHS terms with respect to $\theta$ and $r$. We observe that overall, the thermal-wind balance seems to be  satisfied across the whole shell, and this is more pronounced within the equatorial region.
\par Similarly, in Fig. \ref{fig:TWB}c, we plot the LHS and RHS terms of Eq. (\ref{eq:TWB2}) at all radii from the bottom of the CZ down to  $\sim r_c-\delta_G$  for the anti-solar case with Ro $\approx 0.23$, and we see that the LHS term is somewhat larger than the RHS term within the overshoot region, especially at mid-latitudes. Looking at  Fig. \ref{fig:TWB}d, we find that the LHS term is substantially different from the RHS term especially within the convection zone, thus indicating that the thermal-wind balance is not  satisfied for the anti-solar case in the CZ, although it is somewhat sustained in the bulk of the stable region. We note that Equation (\ref{eq:TWB2}) essentially shows that it is possible to maintain non-cylindrical rotational profiles (i.e. when LHS $\neq 0$) by accounting for latitudinal gradients of $\langle S\rangle$,  but we observe no such behavior in this study.
 
\TWBfig

\subsection{Gyroscopic Pumping}
A well-known mechanism for the generation of meridional circulation via balances in the angular momentum equation, which had been originally extensively studied in the context of the Earth's atmospheric circulation and has been more recently  revisited  in studies of  astrophysical fluid dynamics \citep{McIntyre2007,GA2009,GB2010,Wood2011,MH2011}, is the so-called gyroscopic pumping. More specifically, gyroscopic pumping  is associated with the advection of {\color{black}  angular momentum by}  meridional  {\color{black} flow} due to the divergence of Reynolds stresses and viscous stresses. We now briefly explore the gyroscopic pumping balances in the solar-like case and the anti-solar case to see whether this mechanism takes place in both of these representative runs as well as  which fluxes are dominant  in maintaining it in the convective zone and the overshoot region.  \par Taking the zonal component $\phi$ of the momentum equation and multiplying it by $\lambda=r\sin\theta$, averaging over both time and longitude and finally assuming a steady state, we obtain the conservation of angular momentum  equation of our system given by
\begin{equation}
\label{eq:CAM}
\bar{\rho}\langle \vel_{mc}\rangle\cdot\nabla\mathcal{L}=-(\nabla\cdot \boldsymbol{F}_{RS}+\nabla\cdot \boldsymbol{F}_{VS})
\end{equation}
where $\vel_{mc}=(u_r, u_{\theta})$ and where $\mathcal{L}=\lambda^2\Omega$, which can be non-dimensionally expressed as
\begin{equation}
\mathcal{L}=r\sin\theta\left(\dfrac{r\sin\theta}{\rm Ek}+\langle u_{\phi}\rangle\right).
\end{equation} 
The  non-dimensional transport of angular momentum due to Reynolds stresses and due to viscous stresses is given respectively by
{\color{black}
\begin{equation}
\label{eq:Frs}
\boldsymbol{F}_{RS}=\bar{\rho}r\sin\theta(\langle u_{f,r} u_{f,\phi}\rangle,\langle u_{f,\theta} u_{f,\phi}\rangle),
\end{equation}
\begin{equation}
\label{eq:Fvs}
\boldsymbol{F}_{VS}=\bar{\rho}\left(\langle {u}_{m,\phi}\rangle\sin\theta-r\sin\theta\dfrac{\partial \langle {u}_{m,\phi}\rangle}{\partial r}, \langle {u}_{m,\phi}\rangle\cos\theta-\sin\theta\dfrac{\partial\langle {u}_{m,\phi}\rangle}{\partial\theta}\right).
\end{equation}
}
\GPRo2fig
{\color{black} Following \cite{FM2015}, in} Figure \ref{fig:GPRo2}a, we plot the  left-hand-side term of Equation (\ref{eq:CAM}), i.e. $\bar{\rho}\langle\vel_{mc}\rangle\cdot\nabla\mathcal{L}$ along with the right-hand-side term $-(\nabla\cdot \boldsymbol{F}_{RS}+\nabla\cdot \boldsymbol{F}_{VS})$ at all radii from $r_c$ down to $\sim r_c-\delta_G$ for the solar-like case with Ro $\approx 0.043$ and find that the balance is more or less achieved within the overshoot region  with $\bar{\rho}\langle\vel_{mc}\rangle\cdot\nabla\mathcal{L}$ being slightly larger than $-(\nabla\cdot \boldsymbol{F}_{RS}+\nabla\cdot \boldsymbol{F}_{VS})$. In fact, by looking across the whole shell in Figure \ref{fig:GPRo2}b, we find that, overall, Equation (\ref{eq:CAM}) is satisfied within the CZ and the overshoot region where the differential rotation and the associated meridional flows are generated and advected. We note that we should not expect  Eq. (\ref{eq:CAM}) to be satisfied in the bulk of the radiative zone where the motions are not convectively driven {\color{black} and so are largely axisymmetric}. Furthermore, we notice that $-\nabla\cdot \boldsymbol{F}_{RS}$ seems to be dominant  in the bulk of the CZ while $-\nabla\cdot \boldsymbol{F}_{VS}$ is larger at lower latitudes and the equatorial region, especially close and somewhat below  the bottom of the CZ.
\par In Fig. \ref{fig:GPRo2}c, we show the left-hand-side and the right-hand-side terms of Eq. (\ref{eq:CAM})  at different radii within the overshoot region for the anti-solar case with Ro $\approx 0.23$. We observe that the gyroscopic pumping balance seems to be satisfied within that region although there are some discrepancies between the two terms mostly close to the equator with $-(\nabla\cdot \boldsymbol{F}_{RS}+\nabla\cdot \boldsymbol{F}_{VS})$ being somewhat larger than $\bar{\rho}\langle\vel_{mc}\rangle\cdot\nabla\mathcal{L}$. In Fig. \ref{fig:GPRo2}d, we plot the individual terms across the whole shell and find that the gyroscopic balance is achieved. Similarly to the solar-like case, the divergence of the flux associated with the Reynolds stresses is dominant throughout the shell while the viscous stresses play a less important role overall and are larger mainly at mid-latitudes and the equatorial region of the CZ.
\par Stars operate in a regime where viscosity is very small (with typical Prandtl numbers Pr $\sim O(10^{-6})$), so we would  expect that the advection of the meridional flows would be due to only the divergence of the fluxes associated with the Reynolds stresses while viscous stresses should be negligible, unlike what we see in the runs presented here. Indeed, viscosity seems to have a fairly strong presence near the equatorial region for both the solar-like case and the anti-solar case which would mean that we may need to consider even higher Ra runs (which are computationally much harder to achieve) so that  the viscous stresses would not need to compensate for the weaker Reynolds stresses observed at this Ro and Ra.

\newpage
{\color{black}\section{Discussion}
\label{sec:disc}
\subsection{Summary of our Results and Comparison with Previous Studies of Overshooting Convection}}
We have presented the results of a series of numerical simulations in a two-zone spherical shell in order to examine the effect of rotation and density stratification on the overshooting dynamics associated with the interaction of a  {\color{black} convectively} unstable region  overlying a stably stratified zone. Motivated by solar-type stars, we have assumed a fixed thermal gradient boundary condition at the inner boundary, as well as the existence of an internal heat source whose function  mimics the radiative heating presented in more complicated solar models (Model S), while we have used a fixed aspect ratio which corresponds to  the inner part of  the solar convective region and a large portion of the radiative zone. We have conducted a systematic study by varying the density stratification in the CZ, the Rayleigh number, and the Ekman number while we have assumed a fixed Prandtl number (Pr $=1$). To create our two-layered configuration, we have considered a varying background entropy gradient $d\bar{S}/dr$ which leads to an adiabatic CZ that smoothly transitions into a subadiabatic RZ with a profile fixed for all of our runs -- both non-rotating and rotating.
 
There have been many numerical studies of overshooting convection considering a convective region overlying a stably stratified zone which, as we discussed in Section \ref{sec:intro}, have been mainly focused on the dependence of the overshooting on the relative stability and the transition width between the two layers as well as the degree  of the convective driving in the CZ. In all of these studies, similarly to ours, they found that there is a substantial amount of overshooting $\delta$ below the base of the convective region which depends on the input parameters, the different set-ups and geometries, or whether they are 2D or 3D {\color{black}\citep[see e.g.,][]{Hurlburt86,Hurlburt94,Singh98, Brummell, Rogers2005,Pratt17,Hotta17, Kapyla2019,Korre19}. For instance, \cite{Hurlburt86} reported a value of  $\delta=0.67H_p$, \cite{Singh98} found that $\delta$ ranges between $0.4H_p-0.88H_p$ while  \cite{Brummell} reported values of $\delta$ between $0.4H_p$ and $2H_p$, where $H_p$ is the pressure scale-height. In our runs the pressure scale-height at the bottom of the CZ defined as $H_p=-((1/\bar{P}(r))d\bar{P}(r)/dr)^{-1}|_{r=r_c}$ depends on $N_{\rho}$ and ranges between $\sim 0.32$ and $\sim 46$ such that the overshoot lengthscale in terms of  pressure scale-heights is $0.0074H_p\leq\delta=\delta_G r_o\leq 0.64H_p$. So our findings are similar to those reported in previous studies, especially when we consider higher $N_{\rho}$ cases.}

\par{\color{black} We also found that the computed overshoot lengthscale $\delta_G$ for the non-rotating cases, which have fixed Pr and Ra,  depends on the ratio of the density stratifications in the two zones given by the parameter $R_{\rho}$ such that $\delta_G\propto R_{\rho}^{0.36}$.
However, to the authors' knowledge, there is no previous systematic study on the dependence of overshooting on the density stratification as an input parameter.} A notable exception is the work of \citet{Mass84}, who employed the anelastic approximation in a Cartesian geometry and reported the penetration lengthscale for two different density stratification cases while they also compared these results to a Boussinesq case. In their work, they assumed stacked polytropes with a varying conductivity profile such that a convection zone is sandwiched between two stable layers and  studied upward and downward penetration. However, they did not perform fully non-linear 3D calculations, but instead they  used anelastic modal equations whereby they expanded the horizontal structure of the flow using one- and two-mode hexagonal planforms in order to better resolve the vertical structure.   As a result, their findings have a strong dependence on the chosen horizontal modes and  cannot capture the fully non-linear dynamics associated with overshooting convection.    Nonetheless, they found that the increased density stratification leads to larger penetration and that the anelastic case predicts larger penetration depths than the Boussinesq case. In fact, they reported   strong pure penetration whereby the convection zone extended further down into the stable region and part of the latter became adiabatic.

In our work, we do not observe pure penetration but rather overshooting, although there is a trend in the  $d\tilde{S}_T/dr$ profiles (see Eq. (\ref{eq:dSNrho})) indicating that increasing the stratification along with the Rayleigh number even further might actually result in pure penetration. This can be seen in Fig. \ref{fig:dSRa}, where we plot the total adjusted $d\tilde{S}_T/dr$ profile along with $d\bar{S}/dr$ for the case with $N_{\rho}=3$ at three different Ra. We also notice that the slightly subadiabatic layer around the base of the CZ (discussed in Section \ref{sec:Nrho}) becomes even smaller at larger Ra, where convection is more vigorous.
\dSRafig

 \citet{Mass84} also showed that the local Pe decreased with decreasing stratification which is not the case in our  fully non-linear 3D calculations where increased density stratification leads to smaller Reynolds (and P\'{e}clet numbers) due to somewhat weaker convection at the same Ra for larger $N_{\rho}$ \citep[see e.g.,][]{Jones2009}.   \par We  postulate that in a set-up where $N_{\rho}$ in the CZ increases but somehow the density ratio in the RZ remains the same for all of the cases, then $R_{\rho}$ will be a monotonic function of $N_{\rho}$ and as such it can lead to larger overshoot depths for larger values of $N_{\rho}$, but that remains to be verified by simulations which account for such a  two-layered configuration.

There have been quite a few previous studies that have accounted for rotation in overshooting convective experiments. Using 3D compressible simulations, \citet{Brummell} showed that the overshoot depth decreases with decreasing Ro as a result of  the braking of the vertical flows due to the horizontal interactions of the like-sign vortices as well as the fact that the downflows penetrate at an angle in the presence of strong rotation. They reported a scaling of $\delta\propto$ Ro$^{0.15}$ while they also showed that it depends on the latitude with the poles and the equator having larger $\delta$.
However, their scaling was a result of only three different cases while they also used the {\color{black}$f$}-plane approximation in a Cartesian box which intrinsically cannot account for global flows. Similarly, \citet{Pal2007} found that $\delta$ decreases with Ro with $\delta\propto$ Ro$^{0.2}$ at the poles and $\delta\propto$ Ro$^{0.4}$ at mid-latitudes.  \citet{AugMathis2019} studied overshooting convection under the effect of rotation by building on  the linearized model of \citet{Zahn91}  and they reported  that the overshoot depth decreases like Ro$^{0.3}$.

In this work, we found that $\delta_G\propto$ Ro$^{0.23}$ which is not that different from the predictions of these previous models, even though we operate in a spherical geometry, using a different set-up based on the anelastic approximation. Thus, the decrease in the amount of overshooting with increasing rotation rate seems to be a salient feature of the rotating overshooting dynamics. 
Additionally, we  showed that the overshoot depth depends on latitude with the solar-like case (Ro $\approx 0.043$) leading to larger overshoot depths closer to the poles and at mid-latitudes where the flow is more weakly influenced by rotation and smaller depths near the equator where the convective motions penetrate at an angle due to their alignment with the axis of rotation. For the anti-solar case (Ro $\approx 0.23$), again the overshoot depth is larger at higher-latitudes while it is just a bit larger  close to the equator where the motions are not as much aligned with the axis of rotation   as at mid-latitudes. Possibly this is also a result of the topology of the plumes there which appear to be more like the fly-wheeling motions   described in \citet{Brummell}. Overall when increasing Ro the latitudinal dependence becomes weaker due to the smaller effect of the rotation on convection.

We note that other studies of anelastic overshooting convection in spherical shells \citep[see][]{Browning2004,BMT11,Aug12,Brun2017} showed that the overshoot lengthscale decreases with increasing Rossby number. However, these simulations were not focused on the dependence of the overshooting dynamics on Ro, the measure of the overshoot depth was based on the radial enthalpy flux rather than the full kinetic energy and they were operating at lower Pe. Thus, it is not clear how to draw meaningful comparisons with the results of our study.

{\color{black}\subsection{Implications for Overshooting in the Sun}}
We  showed that for sufficiently small Ro, we can obtain a solar-like differential rotation with a faster equator and slower poles in the CZ transitioning to a uniform rotation in the RZ. However, helioseismology has revealed that the Sun also possesses a conical profile while our solutions lead to cylindrical ones. 
\par Previous studies \citep[e.g.,][]{Rempel2005,Miesch2006}  have shown that  enforcing a strong latitudinal entropy gradient at the base of the CZ  can result in a  conical differential rotation  similarly to what we observe in the Sun.  Looking  at the entropy profile in Fig. 3b of \citet{Miesch2006} (also see Fig.8d from \citet{Matilsky20} who accounted for a fixed flux outer boundary condition instead of a latitudinal entropy gradient at the inner boundary),   we notice that in the solar-like conical profiles achieved in their calculations, the entropy perturbations increase monotonically with latitude away from the equator.

In Figure \ref{fig:Saz}, we show the time- and azimuthally- averaged entropy perturbations  (where we have subtracted the spherically symmetric mean to enhance the latitudinal variations), volume-averaged within the overshoot region ${\langle{S}\rangle}_{ov}$ versus the latitude $\theta$ for the solar-like case of Ro $\approx 0.043$. We observe that   ${\langle{S}\rangle}_{ov}$ is much weaker in magnitude and does not vary substantially in latitude compared with the entropy  profiles corresponding to the conical differential rotation contours in \citet{Miesch2006}. Also, despite the fact that the equatorial region is still cooler than the mid-latitudes and the poles, the profile is non-monotonic. These previous studies  only considered a convection zone and thus they had to enforce a latitudinal entropy gradient at the inner boundary (or assume a fixed flux outer boundary condition) to achieve the tilt of the differential rotation contours. However, in our two-zone  spherical shell model set-up, a strongly varying latitudinal entropy gradient might be obtained self-consistently with a suitable background $d\bar{S}/dr$ profile. Indeed, this could result in stronger entropy gradient perturbations  $\partial \langle S\rangle/\partial\theta$ in the thermally equilibrated state which could in turn lead to non-cylindrical differential rotation profiles. This topic will be addressed in future work.

\Sazfig

{\color{black}
We may attempt to provide some estimates on the solar value of the overshooting lengthscale  using the results of our study. We note, however, that such extrapolations tend to be speculative to some degree as values of many parameters used here are very different from the solar ones. For instance, Ra$_{\odot}\sim \rm O(10^{20})$ and Pr$_{\odot}\sim \rm O(10^{-6})$ \citep[.e.g.,][]{Ossendrijver2003}, while our models operate in a much lower Ra and much higher Pr regime. 

These caveats aside, we can use the scaling relationship between $\delta_G$ and the ratio of the densities $R_{\rho}$ to estimate the solar overshoot depth. We consider the region of the convection zone spanning from the tachocline to the base of the granulation boundary layer ($r_o\approx 0.9983R_{\odot}$),  a region characterized by 11 e-foldings in density \citep[e.g., Model S; ][]{ModelS}.  Using this, and the values of the density at the base of the radiative zone at $r_i\approx 0.2R_{\odot}$ and  at the base of the CZ at $r_c\approx 0.71R_{\odot}$, we find that $R_{\rho}\approx 0.003$.  Based on our scaling of $\delta_G= 0.078R_{\rho}^{0.36}$ from Section \ref{sec:Nrho}, we estimate an overshoot lengthscale for the Sun of $\delta_G \approx 0.0097$ or, equivalently, $\delta=\delta_G r_o\approx 0.1 H_p$, where $H_p\approx 5.7\times 10^9$cm is the pressure scale-height at the base of the solar CZ extracted from Model S.  This estimated value for the solar overshoot lengthscale is similar to the estimates from other studies.  For instance, \cite{Brummell}  used 3D compressible simulations to estimate  an overshoot depth in the range $0.02H_p-0.11H_p$, while  stellar evolution models typically assume overshoot depth values of the order of $0.1H_p$.  Helioseismic studies suggest that the upper limit for the solar overshoot depth is about $0.07H_p$ \citep[see, e.g.][]{Monteiro1994}.

Carrying out a similar exercise based on Rossby number is more difficult as there is still no clear consensus on the value of deep solar convection's Rossby number.  Estimates of convective overturning time based on the B--V color index suggest that Ro$_\odot\sim0.04$ \citep{Corsaro2021}, whereas \cite{Featherstone16b} suggested that a value $\leq 0.01$ was likely based on observed spatial scale of supergranulation.  In principle, this parameter could be inferred for the Sun through helioseismic measurements of deep convection, but attempts to do so have led to inconsistent results \citep[e.g.,][]{Hanasoge,Greer2015,Nagashima2020}. Given these present uncertainties, we refrain from using the scaling law
of $\delta_G$ with Ro to obtain estimates of the solar overshoot
lengthscale.  Alternative observational approaches, however, such as
those based
on inertial mode observations \citep{Gizon2021} may shed further
light on this issue in the future.}

\begin{acknowledgments}
We thank Nic Brummell and Pascale Garaud for useful discussions on overshooting convection over the years.  This work was supported by  NASA grant No. 80NSSC17K0008. LK acknowledges support from the George Ellery Hale Post-Doctoral Fellowship and from NASA's Early Career Investigator Program grant No. 80NSSC21K0455. NF acknowledges funding from NASA grant No. 80NSSC20K0193. The authors acknowledge the NASA High-End-Computing program for providing the computational resources without which the numerical simulations of this work would not have been possible. This work utilized resources from the University of Colorado Boulder Research Computing Group, which is supported by the National Science Foundation (awards ACI-1532235 and ACI-1532236), the University of Colorado Boulder, and Colorado State University.  The Rayleigh code has been developed with support
by the National Science Foundation through the Computational
Infrastructure for Geodynamics  under
 grants NSF-0949446 and NSF-1550901. 

\end{acknowledgments}

\newpage
\appendix
\section{Polytropic Background State in the CZ} 
\label{sec:appendixA}
Following the anelastic benchmark study of \cite{Jones11}, we define a polytropic background state  in the CZ  with
\begin{equation}
\label{eq:polytrope}
\bar{T}=\bar{T}_c z, \quad \bar{\rho}=\bar{\rho}_c z^n, \quad \bar{P}=\bar{P}_c z^{n+1},
\end{equation}
where $n$ is the polytropic index, and where $\bar{T}_c$, $\bar{\rho}_c$, and $\bar{P}_c$ is the value of $\bar{T}$, $\bar{\rho}$ and $\bar{P}$ at the base of the CZ, respectively. 
Assuming that our polytrope satisfies the hydrostatic balance given by
\begin{equation}
\label{eq:hydrobal}
\dfrac{\partial \bar{P}}{\partial r}=-\bar{\rho} {g}=-\dfrac{\bar{\rho}G M}{r^2},
\end{equation}
where $G$ is the gravitational constant and $M$ is the stellar mass
and defining the number of density scale-heights in the CZ given by $N_{\rho}$, we find an expression for the function $z$
\begin{equation}
\label{eq:z}
z=a+\dfrac{b}{r},
\end{equation}  
where 
\begin{equation}
a=\dfrac{f b}{r_{c}}, \quad b=\dfrac{2GM}{5(n+1)c_p\bar{T}_c},
\end{equation}
and where
\begin{equation}
\label{eq:f}
f=\dfrac{({r_{c}}/{r_o}) \exp(N_{{\rho}}/n)-1}{1-\exp(N_{{\rho}}/n)}.
\end{equation}

\bibliography{biblio}
\bibliographystyle{aasjournal}

\end{document}